\documentclass[11pt]{article}

\title{\textbf{Parallel Algorithms for Successive Convolution} \thanks{Last updated on July 22, 2020.
The research of the authors was supported in part by AFOSR grants FA9550-19-1-0281 and FA9550-17-1-0394 and NSF grant DMS 1912183.}}

\author{Andrew J. Christlieb \thanks{Department of Computational Mathematics, Science and Engineering, Michigan State University, East Lansing, MI, 48824, United States; \href{mailto:christli@msu.edu}{christli@msu.edu}.}
\and Pierson T. Guthrey \footnotemark[2] \thanks{Presently at Lawrence Livermore National Laboratory, Livermore, CA, 94550, United States; \href{mailto:piersonguthrey@gmail.com}{guthrey1@llnl.gov}.}
\and William A. Sands \thanks{Department of Computational Mathematics, Science and Engineering, Michigan State University, East Lansing, MI, 48824, United States; \href{mailto:sandswi3@msu.edu}{sandswi3@msu.edu} (corresponding author).}
\and Mathialakan Thavappiragasm \footnotemark[2] \thanks{Presently at Oak Ridge National Laboratory, Oak Ridge, TN, 37830, United States; \href{mailto:mathialakan@gmail.com}{mathialakan@gmail.com}.}
}

\date{}

\usepackage[english]{babel}
\usepackage[utf8]{inputenc}
\usepackage{csquotes}
\usepackage{amsfonts}
\usepackage{amsmath,amssymb, amsthm,accents} 
\usepackage{enumitem}
\usepackage{caption}
\usepackage{subfig}
\usepackage{float}
\usepackage{mathtools}
\usepackage{breqn}
\usepackage[a4paper,margin=0.5in]{geometry}
\usepackage{color}
\usepackage{algorithm}
\usepackage[noend]{algpseudocode}
\usepackage{hyperref}
\hypersetup{
    colorlinks=true,
    linkcolor=blue,
    citecolor=blue,
    filecolor=blue,      
    urlcolor=blue,
}
\usepackage{cleveref}

\usepackage[
backend=biber,
style=ieee,
citestyle=numeric,
sorting=none
]{biblatex}
\usepackage{xcolor}
\usepackage{listings}
 
\definecolor{codegreen}{rgb}{0,0.6,0}
\definecolor{codegray}{rgb}{0.5,0.5,0.5}
\definecolor{codepurple}{rgb}{0.58,0,0.82}
\definecolor{backcolour}{rgb}{0.95,0.95,0.92}
 
\lstdefinestyle{mystyle}{
    backgroundcolor=\color{backcolour},   
    commentstyle=\color{codegreen},
    keywordstyle=\color{magenta},
    numberstyle=\tiny\color{codegray},
    stringstyle=\color{codepurple},
    basicstyle=\ttfamily\footnotesize,
    breakatwhitespace=true,         
    breaklines=true,                 
    captionpos=b,                    
    keepspaces=true,                 
    numbers=left,                    
    numbersep=5pt,                  
    showspaces=false,                
    showstringspaces=false,
    showtabs=false,                  
    tabsize=2
}

\numberwithin{equation}{section}

\makeatletter
\renewcommand*\env@matrix[1][\arraystretch]{%
  \edef\arraystretch{#1}%
  \hskip -\arraycolsep
  \let\@ifnextchar\new@ifnextchar
  \array{*\c@MaxMatrixCols c}}
\makeatother
 
\begin{document}

% Build the title page
\maketitle

% Include each of the sections
\begin{abstract}
The development of modern computing architectures with ever-increasing amounts of parallelism has allowed for the solution of previously intractable problems across a variety of scientific disciplines. Despite these advances, multiscale computing problems continue to pose an incredible challenge to modern architectures because they require resolving scales that often vary by orders of magnitude in both space and time. Such complications have led us to consider alternative discretizations for partial differential equations (PDEs) which use expansions involving integral operators to approximate spatial derivatives \cite{christlieb2019kernel,christlieb2020_NDAD, christlieb2020nonuniformHJE}. These constructions use explicit information within the integral terms, but treat boundary data implicitly, which contributes to the overall speed of the method. This approach is provably unconditionally stable for linear problems and stability has been demonstrated experimentally for nonlinear problems. Additionally, it is matrix-free in the sense that it is not necessary to invert linear systems and iteration is not required for nonlinear terms. Moreover, the scheme employs a fast summation algorithm that yields a method with a computational complexity of $\mathcal{O}(N)$, where $N$ is the number of mesh points along a coordinate direction. While much work has been done to explore the theory behind these methods, their practicality in large scale computing environments is a largely unexplored topic. In this work, we explore the performance of these methods by developing a domain decomposition algorithm suitable for distributed memory systems along with shared memory algorithms. As a first pass, we derive an artificial Courant-Friedrichs-Lewy (CFL) condition that enforces a nearest-neighbor (N-N) communication pattern and briefly discuss possible generalizations. We also analyze several approaches for implementing the parallel algorithms by optimizing predominant loop structures and maximizing data reuse. Using a hybrid design that employs MPI and Kokkos \cite{kokkos2014} for the distributed and shared memory components of the algorithms, respectively, we show that our methods are efficient and can sustain an update rate $> 1\times10^8$ DOF/node/s. We provide results that demonstrate the scalability and versatility of our algorithms using several different PDE test problems, including a nonlinear example, which employs an adaptive time-stepping rule.      
\end{abstract}


% \begin{keywords}
% High-performance computing, performance portability,  domain decomposition, integral solution, fast algorithms, numerical analysis
% \end{keywords}
\section{Introduction}

% What am I presenting here to address the aforementioned problem

In this work, we develop parallel algorithms using novel approaches to represent derivative operators for linear and nonlinear time-dependent partial differential equations (PDEs). We chose to investigate algorithms for these representations due to the stability properties observed for a wide range of linear and nonlinear PDEs. The approach considered here uses expansions involving integral operators to approximate spatial derivatives. Here, we shall refer to this approach as the Method of Lines Transpose (MOL$^T$) though this can be more broadly categorized within a larger class of successive convolution methods. The name arises because the terms in the operator expansions, which we describe later, involve convolution integrals whose operand is recursively or successively defined. Despite the use of explicit data in these integral terms, the boundary data remains implicit, which contributes to both the speed and stability of the representations. The inclusion of more terms in these operator expansions, when combined with a high-order quadrature method, allow one to obtain a high-order discretization in both space and time. Another benefit of this approach is that extensions to multiple spatial dimensions are straightforward as operators can be treated in a line-by-line fashion. Moreover, the integral equations are amenable to fast-summation techniques, which reduce the overall computational complexity, along a given dimension, from $\mathcal{O}(N^2)$ to $\mathcal{O}(N)$, where $N$ is the number of discrete grid points along a dimension. 

High-order successive convolution algorithms have been developed to solve a range of time-dependent PDEs, including the wave equation \cite{causley2014method}, heat equation (e.g., Allen-Cahn \cite{causley2016method} and Cahn-Hilliard equations \cite{causley2017method}), Maxwell's equations \cite{cheng2017asymptotic}, Vlasov equation \cite{christlieb2016weno}, degenerate advection-diffusion (A-D) equation \cite{christlieb2020_NDAD}, and the Hamilton-Jacobi (H-J) equation \cite{christlieb2019kernel, christlieb2020nonuniformHJE}. In contrast to these papers, this work focuses on the performance of the method in parallel computing environments, which is a largely unexplored area of research. Specifically, our work focuses on developing effective domain decomposition strategies for distributed memory systems and building thread-scalable algorithms using the low-order schemes as a baseline. By leveraging the decay properties of the integral representation, we restrict the calculations to localized non-overlapping subsets of the spatial domain. The algorithms presented in this work consider dependencies between nearest-neighbors (N-N), but, as we will see, this restriction can be generalized to include additional information, at the cost of additional communication. Using a hybrid design that employs MPI and Kokkos \cite{kokkos2014} for the distributed and shared memory components of the algorithms, respectively, we show that our methods are efficient and can sustain an update rate $> 1\times10^8$ DOF/node/s. While experimentation on graphics processing units (GPUs) shall be left to future work, we believe choosing Kokkos will provide a path for a more performant and seamless integration of our algorithms with new computing hardware.

Recent developments in successive convolution methods have focused on extensions to solve more general nonlinear PDEs, for which an integral solution is generally not applicable. This work considers discretizations developed for degenerate advection-diffusion (A-D) equations \cite{christlieb2020_NDAD}, as well as the Hamilton-Jacobi (H-J) equations \cite{christlieb2019kernel, christlieb2020nonuniformHJE}. The key idea of these papers exploited the linearity of a given \textit{differential operator} rather than the underlying equations, allowing derivatives in nonlinear problems to be expressed using the same representations developed for linear problems. For linear problems, it was demonstrated that one could couple these representations for the derivative operators with an explicit time-stepping method, such as the strong-stability-preserving Runge-Kutta (SSP-RK) methods, \cite{gottlieb2001strong} and still obtain schemes which maintain unconditional stability \cite{christlieb2019kernel,christlieb2020_NDAD}. To address shock-capturing and control non-physical oscillations, the latter two papers introduced quadratures that use WENO reconstructions, along with a nonlinear filter to further control oscillations. In \cite{christlieb2020nonuniformHJE}, the schemes for the H-J equations were extended to enable calculations on mapped grids. This paper also proposed a new WENO quadrature method that uses a basis that consists of exponential polynomials that improves the shock capturing capabilities. 

% Comparison with other approaches
% How have others approached this problem?
Our choice in discretizing time, first, before treating the spatial quantities, is not a new idea. A well-known approach is Rothe's method \cite{salazar2000theoretical,schemann1998adaptive} in which a finite difference approximation is used for time derivatives and an integral equation solver is developed for the resulting sequence of elliptic PDEs (see e.g., \cite{biros_navier_stokes, biros_stokes_2004,chiu_moore_quaife_transport_2020, quaife_krop_Rothe, quaife_moore_viscous_erosion, huang_stokes_2007, biros_BIE_elliptic_3d}). The earlier developments for successive convolution methods, such as \cite{causley2014method}, are quite similar to Rothe's method in the treatment of the time derivatives. However, successive convolution methods differ from Rothe's method considerably in the treatment of spatial derivatives for nonlinear problems, such as those considered in more recent work on successive convolution (see e.g., \cite{christlieb2019kernel, christlieb2020_NDAD, christlieb2020nonuniformHJE}), as Newton iteration can be avoided on nonlinear terms. Additionally, these methods do not require solutions to linear systems. In contrast, Nystr\"{o}m methods, which are used to discretize the integral equations in Rothe's method, result in dense linear systems, which are typically solved using an iterative method such as GMRES \cite{saad_GMRES_1986}. Despite the fact that the linear systems are well-conditioned, the various collective operations that occur in distributed GMRES solves can become quite expensive on large computing platforms. 

Similarly, in \cite{bruno_lyon_pt_1, bruno_lyon_pt_2}, Bruno and Lyon introduced a spectral method, based on the FFT, for computing spatial derivatives of general, possibly non-periodic, functions known as Fourier-Continuation (FC). They combined this representation with the well-known Alternating-Direction (AD) methods, e.g., \cite{douglasADIheat, douglasADI3space, Peaceman_Rachford}, dubbed FC-AD, to develop implicit solvers suitable for linear equations. This resulted in a method capable of computing spatial derivatives in a rapidly convergent and dispersionless fashion. A domain decomposition technique for the FC method is described in \cite{FC-CNS_pt_1} and weak scaling was demonstrated to 256 processors, using 4 processors per node, but larger runs are not considered. Another related transform approach was developed to solve the linear wave equation \cite{anderson_hybrid_2020}. This work introduced a windowed Fourier methodology and combined this with a frequency-domain boundary integral equation solver to simulate long-time, high-frequency wave propagation. While this particular work does not focus on parallel implementations, they suggest several generic strategies, including a trivially parallelizable approach that solves a collection of frequency-domain integral equations in parallel. However, a purely parallel-in-time approach may not be appropriate for massively parallel systems, especially if few frequencies are required across time. This issue may be further complicated by the parallel implementation of the frequency-domain integral equation solvers, which, as previously mentioned, require the solution of dense linear systems. Therefore, it may be a rather difficult task to develop robust parallel algorithms, which are capable of achieving their peak performance.

% Describe the layout of the paper

This paper is organized as follows: In \cref{sec:Methods}, we provide an overview of the numerical scheme used to formulate our algorithms. To this end, we first illustrate the connections among several characteristically different PDEs through the appearance of common operators in \cref{subsec:Connections}. Using these fundamental operators, we define the integral representation in \cref{subsec:Representation of Derivatives}, which is used to approximate spatial derivatives. Once the representations have been introduced, we briefly discuss the complications associated with boundary conditions and provide the relevant information, in \cref{subsec: Boundary Conditions}, for implementing boundary conditions used in our numerical tests. \Cref{subsec:Spatial Discretization,subsec:Time Integration} briefly review the spatial discretization process (including the fast-summation method and the quadrature) and the coupling with time integration methods, respectively. \Cref{sec:DD Algorithm} provides the details of our new domain decomposition algorithm, beginning with the derivation of the so-called N-N conditions in \cref{subsec: NN criterion}. Using these conditions, we show how this can be used to enforce boundary conditions, locally, for first and second derivative operators (\cref{subsec: NN DD for partial x} and \cref{subsec: NN DD for partial xx}, respectively). Details concerning the implementation of the parallel algorithms are contained entirely in \cref{sec:Implementation}. This includes the introduction of the shared memory programming model (\cref{subsec:Kokkos}), the definition of a certain performance metric (\cref{subsec:Performance metrics}) used in both loop optimization experiments and scaling studies (\cref{subsec: Loop Optimizations}), the presentation of shared memory algorithms (\cref{subsec: Shared Memory Algorithms}), and, lastly, implementation details concerning the distributed memory algorithms (\cref{subsec:Distributed implementation}). \Cref{sec:Results} contains the core numerical results which confirm the convergence (\cref{subsec:Convergence}), as well as, the weak and strong scalability (\cref{subsec:Weak Scaling} and \cref{subsec:Strong Scaling}, respectively) of the proposed algorithms. In \cref{subsec:CFL Studies}, we examine the impact of the restriction posed by the N-N conditions. Finally, we summarize our findings with a brief conclusion in \cref{sec:conclusion}.

\section{Description of Numerical Methods}
\label{sec:Methods}

In this section, we outline the approach used to develop unconditionally stable solvers making use of knowledge for linear operators. We will start by demonstrating the connections between several different PDEs using operator notation, which will allow us to reuse, or combine, approximations in several different ways. Once we have established these connections, we define an appropriate ``inverse" operator and use this to develop the expansions used to represent derivatives. The representations we develop for derivative operators are motivated by the solution of simple 1-D problems. However, in multi-dimensional problems, these expressions are still valid approximations, in a certain sense, even though the kernels in the integral representation may not be solutions to the PDE in question. While these approximations can be made high-order in both time and space, the focus of this work is strictly on the scalability of the method, so we will limit ourselves to formulations which are first-order in time. Note that the approach described in \cref{sec:DD Algorithm,sec:Implementation}, which considers first-order schemes, is quite general and can be easily extended for high-order representations. Once we have discussed our treatment of derivative terms, we describe the fast summation algorithm and quadrature method in \cref{subsec:Spatial Discretization}. Despite the fact that this work only considers smooth test problems, we include the relevant modifications required for non-smooth problems for completeness. In \cref{subsec:Time Integration}, we illustrate how the representation of derivative operators can be used within a time stepping method to solve PDEs. 

\subsection{Connections Among Different PDEs}
\label{subsec:Connections}

Before introducing the operators relevant to successive convolution algorithms, we establish the operator connections appearing in several linear PDE examples. This process helps identify key operators that can be represented with successive convolution. Specifically, we shall consider the following three prototypical linear PDEs:
\begin{itemize}[topsep=1em,itemsep=1em]
% \begin{itemize}
    \item Linear advection equation: $(\partial_t -c\partial_x)u=0$,
    \item Diffusion equation: $(\partial_t -\nu \partial_{xx})u=0$,
    \item Wave equation: $(\partial_{tt} -c^2 \partial_{xx})u=0$.
\end{itemize}
Next, we apply an \textit{implicit} time discretization to each of these problems. For discussion purposes, we shall consider lower-order time discretizations, i.e., backward-Euler for the $\partial_t u$ and a second-order central difference for $\partial_{tt}u$. If we identify the current time as $t^n$, the new time level as $t^{n+1}$, and $\Delta t = t^{n+1} - t^n$, then we obtain the corresponding set of semi-discrete equations:
\begin{itemize}[topsep=1em,itemsep=1em]
% \begin{itemize}
    \item Linear advection equation: $(\mathcal{I} -\Delta t c\partial_x)u^{n+1}=u^{n}$,
    \item Diffusion equation: $(\mathcal{I} -\Delta t \nu \partial_{xx})u^{n+1}=u^{n}$,
    \item Wave equation: $(\mathcal{I} -\Delta t^2 c^2 \partial_{xx})u^{n+1}=2u^{n}-u^{n-1}$.
\end{itemize}
Here, we use $\mathcal{I}$ to denote the identity operator, and, in all cases, each of the spatial derivatives are taken at time level $t^{n+1}$ to keep the schemes implicit. The key observation is that the operator $(\mathcal{I} \pm\frac{1}{\alpha} \partial_x)$ arises in each of these examples. Notice that $$(\mathcal{I} -\frac{1}{\alpha^2} \partial_{xx}) = (\mathcal{I}-\frac{1}{\alpha} \partial_x)(\mathcal{I}+\frac{1}{\alpha} \partial_x) ,$$ where $\alpha$ is a parameter that is selected according to the equation one wishes to solve. For example, in the case of diffusion, one selects $$ \alpha = \frac{\beta}{\sqrt{\nu \Delta t }}, $$ while for the linear advection and wave equations, one selects $$ \alpha = \frac{\beta}{c\Delta t }.$$  The parameter $\beta$, which does not depend on $\Delta t$, is then used to tune the stability of the approximations. For test problems appearing in this paper, we always use $\beta = 1$. In \cref{subsec:Representation of Derivatives}, we demonstrate how the operator $(\mathcal{I} \pm\frac{1}{\alpha} \partial_x)$ can be used to approximate spatial derivatives. We remark that for second derivatives, one can also use $(\mathcal{I} -\frac{1}{\alpha^2} \partial_{xx})$ to obtain a representation for second order spatial derivatives, instead of factoring into ``left" and ``right" characteristics. 

Next, we introduce the following definitions to simplify the notation:
\begin{equation}
    \label{eq:left and right first derivative ops}
    \mathcal{L}_{L} \equiv \mathcal{I} - \frac{1}{\alpha} \partial_x, \quad \mathcal{L}_{R} \equiv \mathcal{I} + \frac{1}{\alpha} \partial_x.
\end{equation}
Written in this manner, these definitions indicate the left and right-moving components of the characteristics, respectively, as the subscripts are associated with the direction of propagation. For second derivative operators, which are not factored into first derivatives, we shall use
\begin{equation}
    \label{eq:second derivative op}
    \mathcal{L}_{0} \equiv \mathcal{I} - \frac{1}{\alpha^2} \partial_{xx}.
\end{equation}

In order to connect these operators with suitable expressions for spatial derivatives, we need to define the corresponding ``inverse" for each of these \textit{linear} operators on a 1-D interval $[a,b]$. These definitions are given as
\begin{align}
    \label{eq:left-moving inverse op first derivative}
    \mathcal{L}_L^{-1}[~\cdot~; \alpha](x) &\equiv \alpha \int_x^b e^{-\alpha(s-x)}(\cdot) \,ds + B e^{-\alpha(b-x)}, \\
        &\equiv I_L[~\cdot~; \alpha](x) + B e^{-\alpha(b-x)}, \nonumber \\
    \label{eq:right-moving inverse op first derivative}
    \mathcal{L}_R^{-1}[~\cdot~; \alpha](x) &\equiv \alpha \int_a^x e^{-\alpha(x-s)}(\cdot) \,ds + A e^{-\alpha(x-a)}, \\
        &\equiv I_R[~\cdot~; \alpha](x) + A e^{-\alpha(x-a)}. \nonumber
\end{align}
These definitions can be derived in a number of ways. In \cref{subsec:MOLT for Advection}, we demonstrate how these definitions can be derived for the linear advection equation using the integrating factor method. In these definitions, $A$ and $B$ are constants associated with the ``homogeneous solution" of a corresponding semi-discrete problem and are used to satisfy the boundary conditions. In a similar way, one can compute the inverse operator for definition \eqref{eq:second derivative op}, which yields
\begin{align}
    \label{eq:second derivative inverse op}
    \mathcal{L}_0^{-1}[~\cdot~; \alpha](x) &\equiv \frac{\alpha}{2} \int_a^b e^{-\alpha | x - s |}(\cdot) \,ds + A e^{-\alpha(x-a)} + B e^{-\alpha(b-x)}, \\
                &\equiv  I_0[~\cdot~; \alpha](x) + A e^{-\alpha(x-a)} + B e^{-\alpha(b-x)}. \nonumber
\end{align}
In these definitions, we refer to ``$\cdot$" as the operand and, again, $\alpha$ is a parameter selected according to the problem being solved. Although it is a slight abuse of notation, when it is not necessary to explicitly indicate the parameter or the point of evaluation, we shall place the operand inside a pair of parenthesis.

If we connect these definitions to each of the linear semi-discrete equations mentioned earlier, we can determine the update equation through an \textit{analytic inversion} of the corresponding linear operator(s): 
\begin{itemize}[topsep=1em,itemsep=1em]
%\begin{itemize}
    \item Linear advection equation: $u^{n+1} =\mathcal{L}_R^{-1}(u^{n})$, or $u^{n+1} = \mathcal{L}_L^{-1}(u^{n})$,
    \item Diffusion equation: $u^{n+1} =\mathcal{L}_L^{-1}(\mathcal{L}_R^{-1}(u^{n}))$, or $u^{n+1} = \mathcal{L}_{0}^{-1}(u^{n})$,
    \item Wave equation: $u^{n+1} =\mathcal{L}_L^{-1}(\mathcal{L}_R^{-1}(2u^{n}-u^{n-1}))$, or $u^{n+1} = \mathcal{L}_{0}^{-1}(2u^{n}-u^{n-1})$,
\end{itemize}
with the appropriate choice of $\alpha$ for the problem being considered. We note that each of these methods can be made high-order following the work in \cite{christlieb2019kernel,  christlieb2020_NDAD, christlieb2020nonuniformHJE,causley2016method, causley2014higher}, where it was demonstrated that these approaches lead to methods that are unconditionally stable to all orders of accuracy for these linear PDEs, even with variable wave speeds or diffusion coefficients. Since the process of analytic inversion yields an integral term, a fast-summation technique should be used to reduce the computational complexity of a naive implementation, which would otherwise scale as $\mathcal{O}(N^2)$. Some details concerning the spatial discretization and the $\mathcal{O}(N)$ fast-summation method are briefly summarized in \cref{subsec:Spatial Discretization} (for full details, please see \cite{causley2016method}). Next we demonstrate how the operator $\mathcal{L}_{*}$ can be used to approximate spatial derivatives.

\subsection{Representation of Derivatives}
\label{subsec:Representation of Derivatives}

In the previous section, we observed that characteristically different PDEs can be described in-terms of a common set of operators. The focus of this section shall be on manipulating these approximations to obtain a high-order discretization in time through certain operator expansions. The process begins by introducing an operator related to $\mathcal{L}_{*}^{-1}$, namely,
\begin{equation}
    \label{eq:D operators}
    \mathcal{D}_{*} \equiv \mathcal{I} - \mathcal{L}_{*}^{-1},
\end{equation}
where $*$ can be $L$, $R$, or $0$. The motivation for these definitions will become clear soon. Additionally, we can derive an identity from the definitions \eqref{eq:D operators}. By manipulating the terms we quickly find that
\begin{equation}
    \label{eq:D operator identity with inverse}
    \mathcal{L}_{*} \equiv \left( \mathcal{I} - \mathcal{D}_{*} \right)^{-1},
\end{equation}
again, where $*$ can be $L$, $R$, or $0$. The purpose of the identity \eqref{eq:D operator identity with inverse} is that it connects the spatial derivative to an expression involving integrals of the solution rather than derivatives. In other words, it allows us to avoid having to use a stencil operation for derivatives. 

To obtain an approximation for the first derivative in space, we can use $\mathcal{L}_{L}$, $\mathcal{L}_{R}$, or both of them, which may occur as part of a monotone splitting. If we combine the definition of the left propagating first derivative operator in equation \eqref{eq:left and right first derivative ops} with the definition \eqref{eq:D operators} and identity \eqref{eq:D operator identity with inverse}, we can define the first derivative in terms of the $\mathcal{D}_{L}$ operator. Observe that
\begin{align}
    \partial_{x}^{+} &= \alpha \left( \mathcal{I} - \mathcal{L}_{L} \right), \nonumber \\*
                     &= \alpha \left( \mathcal{L}_{L}\mathcal{L}_{L}^{-1} - \mathcal{L}_{L} \right), \nonumber \\*
                     &= \alpha \mathcal{L}_{L} \left( \mathcal{L}_{L}^{-1} - \mathcal{I} \right), \nonumber \\*
                     &= -\alpha \mathcal{L}_{L} \left( \mathcal{I} - \mathcal{L}_{L}^{-1} \right), \nonumber \\*
                     &= -\alpha \left( \mathcal{I} - \mathcal{D}_{L} \right)^{-1} \mathcal{D}_{L}, \nonumber \\*
                     &= -\alpha \sum_{p=1}^{\infty} \mathcal{D}_{L}^{p}, \label{eq:right-biased first derivative}
\end{align}
where, in the last step, we used the fact that the operator $\mathcal{D}_{L}$ is bounded by unity in an operator norm. We use the $+$ convention to indicate that this is a right-sided derivative. Likewise, for the right propagating first derivative operator, we find that the complementary left-biased derivative is given by
\begin{equation}
    \label{eq:left-biased first derivative}
     \partial_{x}^{-} = \alpha \sum_{p=1}^{\infty} \mathcal{D}_{R}^{p}.
\end{equation}
Additionally, the second derivative can be expressed as
\begin{equation}
    \label{eq:representation of second derivative in D}
     \partial_{xx} = -\alpha^2 \sum_{p=1}^{\infty} \mathcal{D}_{0}^{p}.
\end{equation}
As the name implies, each power of $\mathcal{D}_{*}$ is \textit{successively} defined according to
\begin{equation}
    \label{eq:successive relation for D}
    \mathcal{D}_{*}^{k} \equiv \mathcal{D}_{*} \left( \mathcal{D}_{*}^{k-1} \right).
\end{equation}

In previous work, \cite{christlieb2020_NDAD}, for periodic boundary conditions, it was established that the partial sums for the left and right-biased approximations to $\partial_x$ satisfy 
\begin{equation}
    \label{eq:Truncation Thm first derivatives}
    \partial_x^+ =  -\alpha \left ( \sum_{p=1}^{n} \mathcal{D}_L^p + \mathcal{O}\left ( \frac{1}{\alpha^{n+1}}\right )\right ), \quad
    \partial_x^- = \alpha \left ( \sum_{p=1}^{n} \mathcal{D}_R^p + \mathcal{O}\left ( \frac{1}{\alpha^{n+1}}\right ) \right ).
\end{equation}
Similarly for second derivatives, with periodic boundaries, retaining $n$ terms leads to a truncation error with the form 
\begin{equation}
    \label{eq:Truncation Thm for second derivatives}
    \partial_{xx} = -\alpha^2 \left( \sum_{p=1}^{n} \mathcal{D}_{0}^p + \mathcal{O}\left ( \frac{1}{\alpha^{2n+2}}\right )  \right).
\end{equation}
In both cases, the relations can be obtained through a repeated application of integration by parts with induction. These approximations are still exact, in space, but the integral operators nested in $\mathcal{D}_{*}$ will eventually be approximated with quadrature. From these relations, we can also observe the impact of $\alpha$ on the size of the error in time. In particular, if we select $\alpha = \mathcal{O} \left( 1/\Delta t \right)$ in \eqref{eq:Truncation Thm first derivatives} and $\alpha = \mathcal{O} \left( 1/\sqrt{\Delta t} \right)$ in \eqref{eq:Truncation Thm for second derivatives}, each of the approximations should have an error of the form $\mathcal{O}(\Delta t^n)$. The results concerning the consistency and stability of these higher order approximations were established in \cite{christlieb2019kernel, christlieb2020_NDAD,causley2014higher}.

As mentioned earlier, this work only considers approximations which are first-order with respect to time. Therefore, we shall restrict ourselves to the following operator representations:
\begin{equation}
    \label{eq:representations used in paper}
    \partial_x^+ \approx -\alpha \mathcal{D}_L, \quad \partial_x^- \approx \alpha \mathcal{D}_R, \quad \partial_{xx} \approx -\alpha^{2} \mathcal{D}_{0}. 
\end{equation}
This is a consequence of retaining a single term from each of the partial sums in equations \eqref{eq:right-biased first derivative}, \eqref{eq:left-biased first derivative}, and \eqref{eq:representation of second derivative in D}. Consequently, computing higher powers of $\mathcal{D}_{*}$ is unnecessary, so the successive property \eqref{eq:successive relation for D} is not needed. However, it indicates, clearly, a possible path for higher-order extensions of the ideas which will be presented here. For the moment, we shall delay prescribing the choice of $\alpha$ used in the representations \eqref{eq:representations used in paper}, in order to avoid a problem-dependent selection. As alluded to at the beginning of this section, an identical representation is used for multi-dimensional problems, where the $\mathcal{D}_{*}$ operators are now associated with a particular dimension of the problem. The operators along a particular dimension are constructed using data along that dimension of the domain, so that each of the directions remains uncoupled. This completes the discussion on the generic form of the representations used for spatial derivatives. Next, we provide some information regarding the treatment of boundary conditions, which determine the constants $A$ and $B$ appearing in the $\mathcal{D}_{*}$ operators.    

\subsection{Comment on Boundary Conditions}
\label{subsec: Boundary Conditions}

The process of prescribing values of $A$ and $B$, inside equations \cref{eq:left-moving inverse op first derivative,eq:right-moving inverse op first derivative,eq:second derivative inverse op}, which are required to construct $\mathcal{D}_{*}$, is highly dependent on the structure of the problem being solved. Previous work has shown how to prescribe a variety of boundary conditions for linear PDEs (see e.g., \cite{causley2014method, causley2016method,causley2017method,causley2014higher}). For example, in linear problems, such as the wave equation, with either periodic or non-periodic boundary conditions, one can directly enforce the boundary conditions to determine the constants $A$ and $B$. The situation can become much more complicated for problems which are both nonlinear and non-periodic. For approximations of at most third-order accuracy, with nonlinear PDEs, one can use the techniques from \cite{christlieb2019kernel} for non-periodic problems. To achieve high-order time accuracy, the partial sums for this case were modified to eliminate certain low-order terms along the boundaries.  We note that the development of high-order time discretizations, subject to non-trivial boundary conditions, for nonlinear operators, is still an open area of research for successive convolution methods. 

As this paper concerns the scalability of the method, we shall consider test problems that involve periodic boundary conditions. For periodic problems defined on the line interval $[a,b]$, the constants associated with the boundary conditions for first derivatives are given by
\begin{equation}
    \label{eq:constants A and B for partial x}
    A = \frac{I_R [v; \alpha](b)}{1 - \mu}, \quad B = \frac{I_L [v; \alpha](a)}{1 - \mu},
\end{equation}
where $I_L$ and $I_R$ were defined in equations \eqref{eq:left-moving inverse op first derivative} and \eqref{eq:right-moving inverse op first derivative}. Similarly, for second derivatives, the constants can be determined to be
\begin{equation}
    \label{eq:constants A and B for partial xx}
    A = \frac{I_0 [v; \alpha](b)}{1 - \mu}, \quad B = \frac{I_0 [v; \alpha](a)}{1 - \mu},
\end{equation}
with the definition of $I_0$ coming from \eqref{eq:second derivative inverse op}. In the expressions \eqref{eq:constants A and B for partial x} and \eqref{eq:constants A and B for partial xx} provided above, we use
\begin{equation*}
    \mu \equiv e^{-\alpha (b-a)},
\end{equation*}
where $\alpha$ is the appropriately chosen parameter. Note that the function $v(x)$ denotes the generic operand of the operator $\mathcal{D}_{*}$. This helps reduce the complexity of the notation when several applications of $\mathcal{D}_{*}$ are required, since they are recursively defined. As an example, suppose we wish to compute $\partial_{xx} h(u)$, where $h$ is some known function. For this, we can use the first-order scheme for second derivatives (see \eqref{eq:representations used in paper}) and take $v = h(u)$ in the expressions \eqref{eq:constants A and B for partial xx} for the boundary terms.

\subsection{Fast Convolution Algorithm and Spatial Discretization}
\label{subsec:Spatial Discretization}

\begin{figure}[t]
    \centering
    \includegraphics[width=0.5\linewidth]{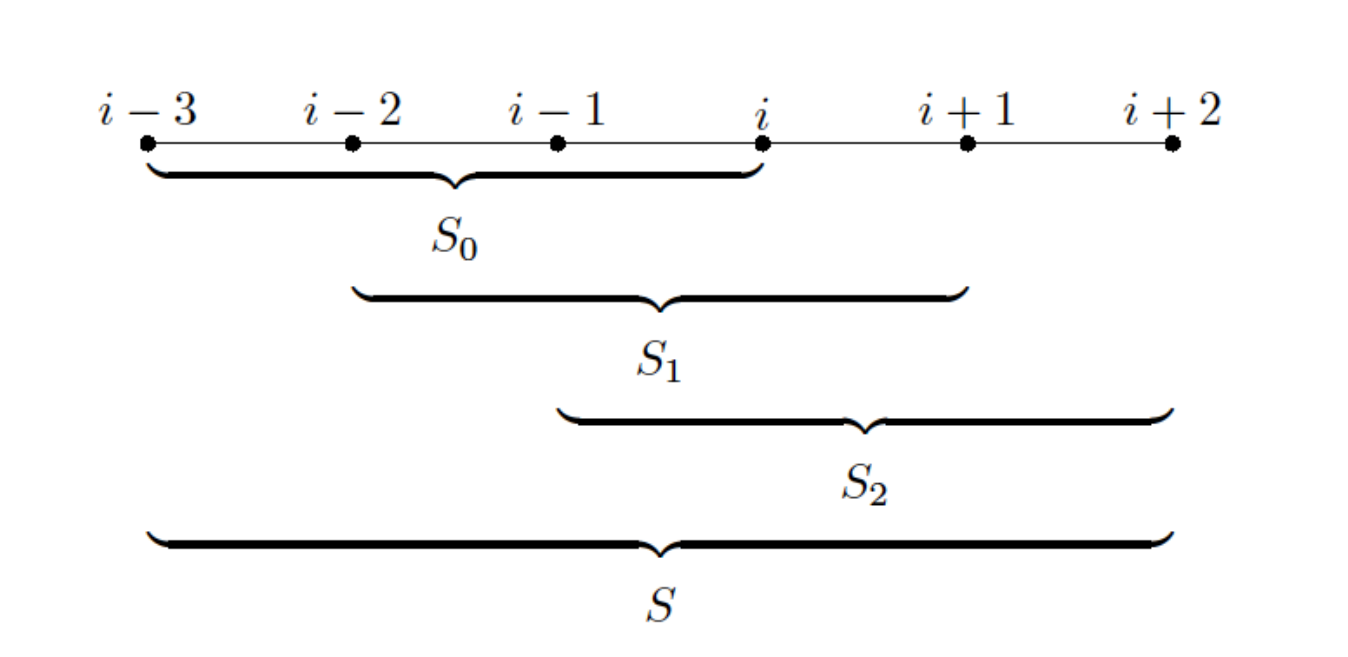}
    \caption{Stencils used to build the sixth-order quadrature \cite{christlieb2019kernel,christlieb2020_NDAD}}
    \label{fig:WENO stencil}
\end{figure}

To perform a spatial discretization over $[a,b]$, we first create a grid of $N + 1$ points:
\begin{equation*}
    x_{i} = a + i\Delta x_{i}, \quad i = 0, \cdots, N,
\end{equation*}
where 
\begin{equation*}
    \Delta x_{i} = x_{i+1} - x_{i}.
\end{equation*}
A naive approach to computing the convolution integral would lead to method of complexity $\mathcal{O}(N^{2})$, where $N$ is the number of grid points. However, using some algebra, we can write recurrence relations for the integral terms which comprise $\mathcal{D}_{*}$:
\begin{align*}
    &I_{R}[v; \alpha](x_{i}) = e^{-\alpha \Delta x_{i-1}} I_{R}[v; \alpha](x_{i-1}) +  J_{R}[v; \alpha](x_{i}), \quad I_{R}[v; \alpha](x_{0}) = 0, \\
    &I_{L}[v; \alpha](x_{i}) = e^{-\alpha \Delta x_{i}} I_{L}[v; \alpha](x_{i+1}) + J_{L}[v; \alpha](x_{i}), \quad I_{L}[v; \alpha](x_{N}) = 0.
\end{align*}
Here, we have defined the local integrals
\begin{align}
    \label{eq:JR}
    J_{R}[v; \alpha](x_{i}) &= \alpha \int_{x_{i-1}}^{x_{i}} e^{-\alpha(x_{i} - s)} v(s) \,ds, \\
    \label{eq:JL}
    J_{L}[v; \alpha](x_{i}) &= \alpha \int_{x_{i}}^{x_{i+1}} e^{-\alpha(s-x_{i})} v(s) \,ds. 
\end{align}
By writing the convolution integrals this way, we obtain a summation method which has a complexity of $\mathcal{O}(N)$. Note that the same algorithm can be applied to compute the convolution integral for the second derivative operator by splitting the integral at a point $x$. After applying the above algorithm to the left and right contributions, we can recombine them through ``averaging" to recover the original integral. While a variety of quadrature methods have been proposed to compute the local integrals \eqref{eq:JL} and \eqref{eq:JR} (see e.g., \cite{christlieb2020nonuniformHJE, causley2014method,causley2016method, causley2017method, causley2014higher, causley2013method}), we shall consider sixth-order quadrature methods introduced in \cite{christlieb2020_NDAD}, which use WENO interpolation to address both smooth and non-smooth problems. In what follows, we describe the procedure for $J_{R}[v; \alpha](x_{i})$, since the reconstruction for $J_{L}[v; \alpha](x_{i})$ is similar. This approximation uses a six-point stencil given by
\begin{equation*}
    S(i) = \{ x_{i-3}, \cdots, x_{i+2} \},
\end{equation*}
which is then divided into three smaller stencils, each of which contains four points, defined by $S_{r} = \{ x_{i-3+r}, \cdots, x_{i+r} \}$ for $r = 0,1,2$. We associate $r$ with the shift in the stencil. A graphical depiction of this stencil is provided in \cref{fig:WENO stencil}. The quadrature method is developed as follows:
\begin{enumerate}[topsep=1em,itemsep=1em]
% \begin{enumerate}
    \item On each of the small stencils $S_{r}(i)$., we use the approximation
    \begin{equation}
        \label{eq:Approx on smaller stencil}
        J_{R}^{(r)}[v; \alpha](x_{i}) \approx \alpha \int_{x_{i-1}}^{x_{i}} e^{-\alpha(x_{i} - s)} p_{r}(s) \,ds = \sum_{j = 0}^{3} c_{-3 + r + j}^{(r)} v_{-3 + r + j},
    \end{equation}
    where $p_{r}(x)$ is the Lagrange interpolating polynomial formed from points in $S_{r}(i)$ and $c_{\ell}^{(r)}$ are the interpolation coefficients, which depend on the parameter $\alpha$ and the grid spacing, but not $v$.
    \item In a similar way, on the large stencil $S(i)$ we obtain the approximation
    \begin{equation}
        \label{eq:Approx on big stencil}
        J_{R}[v; \alpha](x_{i}) \approx \alpha \int_{x_{i-1}}^{x_{i}} e^{-\alpha(x_{i} - s)} p(s) \,ds.
    \end{equation}
    \item When function $v(x)$ is smooth, we can combine the interpolants on the smaller stencils, so they are consistent with the high-order approximation obtained on the larger stencil, i.e.,
    \begin{equation}
        \label{eq:WENO linear weights}
        J_{R}[v; \alpha](x_{i}) = \sum_{r = 0}^{2} d_{r}J_{R}^{(r)}[v; \alpha](x_{i}),
    \end{equation}
    where $d_{r} > 0$ are called the linear weights, which form a partition of unity. The problems we consider in this work involve smooth functions, so this is sufficient for the final approximation. For instances in which the solution is not smooth, the linear weights can be mapped to nonlinear weights using the notion of smoothness. We refer the interested reader to our previous work \cite{christlieb2019kernel, christlieb2020_NDAD, christlieb2020nonuniformHJE} for details concerning non-smooth data sets.
    % \item When the function $v(x)$ is not smooth, we can use the notion of ``smoothness" to eliminate oscillatory substencils in the collocation. Here, the smoothness of $v(x)$ on the interpolation stencil $S_{r}$, denoted as $\beta_r$, can be defined using a variety of methods, such as undivided differences \cite{christlieb2020nonuniformHJE}. 
    % % \begin{equation}
    % %     \label{eq:smoothness indicator def}
    % %     \beta_{r} = \sum_{\ell = 2}^{\ell = 3} \int_{x_{i-1}}^{x_{i}} \Delta x_{i}^{2\ell - 3} \left( \frac{\partial^{\ell} p_{r}(x)}{\partial x^{\ell}} \right)^{2} \, dx.
    % % \end{equation}
    % \item Next the smoothness indicators are used to map the linear weights to the nonlinear weights. For example, authors in \cite{christlieb2020_NDAD,christlieb2019kernel, christlieb2020nonuniformHJE} employed the WENO-Z scheme \cite{borges2008improved}, which uses the mapping
    % \begin{equation}
    %     \label{eq:WENO-Z nonlinear weights}
    %     \widetilde{\omega}_{r} = d_{r}\left(1 + \frac{\tau}{\epsilon + \beta_{r}} \right), \quad \tau = \lvert \beta_{0} - \beta_{2} \rvert, \quad \epsilon = 10^{-6}.
    % \end{equation}
    % The result is then rescaled as
    % \begin{equation*}
    %     \omega_{r} = \widetilde{\omega}_{r}/\sum_{s = 0}^{2} \widetilde{\omega}_{s}
    % \end{equation*}
    % to enforce the condition that the $\omega_{r}$'s sum to unity. 
    % \item As part of their development of a high-order scheme, authors in \cite{christlieb2020_NDAD,christlieb2019kernel} devised a nonlinear filter to limit high-order corrections. We do not consider this development here as our work uses first-order schemes. 
\end{enumerate}
In \cref{sec:WENO quad info}, we provide the expressions used to compute coefficients $c_{\ell}^{(r)} \text{ and } d_{r}$ for a uniform grid, although the non-uniform grid case can be done as well \cite{shu2009high}. In the case of a non-uniform mesh, the linear weights $d_r$ would become locally defined in the neighborhood of a given point and would need to be computed on-the-fly. Uniform grids eliminate this requirement as the linear weights, for a given direction, can be computed once per time step and reused in each of the lines pointing along that direction. 

% Replace this with the description from the Sandia report
\subsection{Coupling Approximations with Time Integration Methods}
\label{subsec:Time Integration}

Here, we demonstrate how one can use this approach to solve a large class of PDEs by coupling the spatial discretizations in \cref{subsec:Spatial Discretization} with explicit time stepping methods. In what follows, we shall consider general PDEs of the form
\begin{equation*}
    \partial_t U = F(t,U),
\end{equation*}
where $F(t,U)$ is a collective term for spatial derivatives involving the solution variable $U$. Possible choices for $F$ might include generic nonlinear advection and diffusion terms
\begin{equation*}
    F(t,U) = \partial_x g_1(U)+ \partial_{xx} g_2(U),
\end{equation*}
or even components of the HJ equations
\begin{equation*}
    F(t,U) = H(U, \partial_x U).
\end{equation*}

To demonstrate how one can couple these approaches, we start by discretizing a PDE in time, but, rather than use backwards Euler, we use an $s$-stage explicit Runge-Kutta (RK) method, i.e.,
$$
u_{n+1} = u_{n}+\sum_{i=1}^s b_i k_i,
$$
where the various stages are given by
\begin{align*}
    k_1 &= F(t^{n}, u^{n}), \\
    k_2 &= F(t^{n} + c_2\Delta t, u^{n}+\Delta t a_{21} k_1),\\
    & \vdots \\
    k_s &= F\left (t^{n}+c_s\Delta t, u^{n}+\Delta t\sum_{j=1}^s a_{sj}k_j \right ).
\end{align*}
As with a standard Method-of-Lines (MOL) discretization, we would need to reconstruct derivatives within each RK-stage. To illustrate, consider the nonlinear A-D equation, $$F(t,u)= \partial_x g_1(u)+ \partial_{xx} g_2(u).$$ For a term such as $g_1$, we would use a monotone Lax-Friedrichs flux splitting, i.e., $g_1\sim \frac{1}{2}(g_1^+ + g_1^-)$, where $g_1^\pm = \frac{1}{2} ( g_1(u) \pm r u)$ with $r=\max_u g_1'(u) $.  Hence, a particular RK-stage can be approximated using
\begin{equation*}
    F(t,u) \approx  -\frac{1}{2} \alpha \sum_{p=1}^{s} \mathcal{D}_L^p [ g_1^+(u) ; \alpha] + \frac{1}{2} \alpha \sum_{p=1}^{s} \mathcal{D}_R^p [ g_1^-(u) ; \alpha] + \alpha_{\nu}^{2} \sum_{p=1}^{s} \mathcal{D}_0^p [g_2(u) ; \alpha_{\nu} ].
\end{equation*}
The resulting approximation to the RK-stage can be shown to be $\mathcal{O}(\Delta t^{s+1})$ accurate. Another nonlinear PDE of interest to us is the H-J equation $$F(t,u)=H(\partial_x u ).$$ In a similar way, we would replace the Hamiltonian with a monotone numerical Hamiltonian, such as
\begin{equation*}
    \hat{H}(v^-, v^+) = H\left (\frac{v^-+v^+}{2}\right) + r(v^-,v^+)\frac{v^- - v^+}{2},
\end{equation*}
where $r(v^-,v^+)= \max_v H'(v)$. Then, the left and right derivative operators in the numerical Hamiltonian can be replaced with $$ \partial_x^- u = \alpha \sum_{p=1}^{s} \mathcal{D}_R^p [u; \alpha] , \quad \partial_x^+ u = -\alpha \sum_{p=1}^{s} \mathcal{D}_L^p [u; \alpha],  $$ which, again, yields an $\mathcal{O}(\Delta t^{s+1})$ approximation. 

In previous work \cite{christlieb2019kernel,christlieb2020_NDAD}, for linear forms of $F(t,u)$, it was shown that the resulting methods are unconditionally stable when coupled to explicit RK methods, up to order 3. Extensions beyond third-order are, indeed, possible, but were not considered. For general, nonlinear problems, we typically couple an $s$-stage RK method to a successive convolution approximation of the same time accuracy, so that the error in the resulting approximation is $\mathcal{O}(\Delta t^{s})$.

\section{Nearest-Neighbor Domain Decomposition Algorithm}
\label{sec:DD Algorithm}

% see https://arxiv.org/pdf/1306.6902.pdf

In this section, we provide the relevant mathematical definitions of our domain decomposition algorithm, which are derived from the key operators used in successive convolution. Our goal is to establish and exploit data locality in the method, so that certain reconstructions, which are nonlocal, can be independently completed on non-overlapping blocks of the domain. This is achieved in part by leveraging certain decay properties of the integral representations. Once we have established some useful definitions, we use them to derive conditions in \cref{subsec: NN criterion}, which restrict the communication pattern to N-Ns. Then in \cref{subsec: NN DD for partial x,subsec: NN DD for partial xx}, we illustrate how this condition can be used to enforce boundary conditions, in a consistent manner, for first and second derivative operators, on each of the blocks. We then provide a brief summary of these findings along with additional comments in \cref{subsec: Sec 3 Additional Comments}.

\begin{figure}[t]
    \centering
        \centering
        \includegraphics[width=0.7\linewidth]{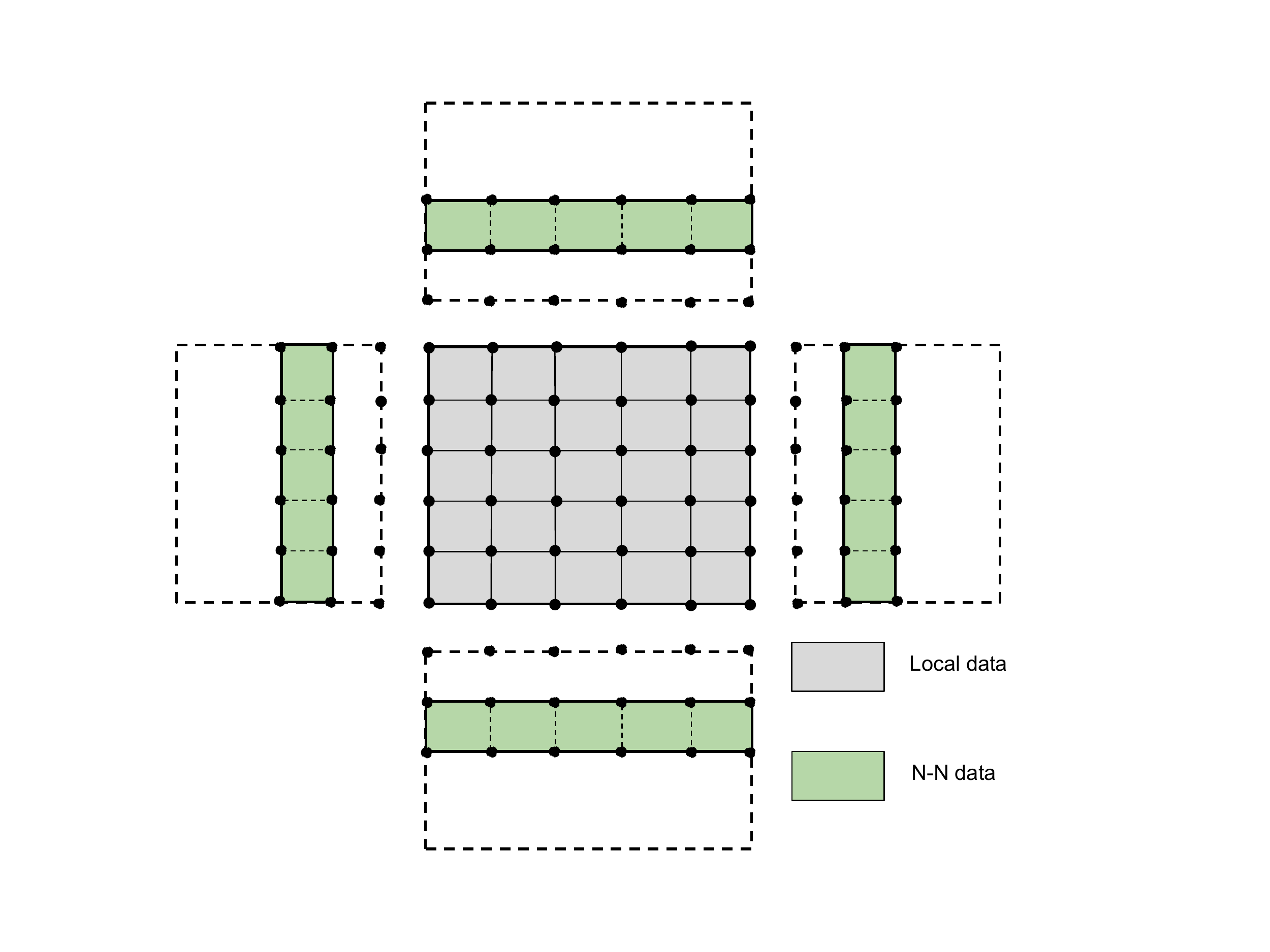}
        \caption{A sixth-order WENO quadrature stencil in 2-D.}
        \label{fig:quad_DD}
\end{figure}

Maintaining a localized stencil is often advantageous in parallel computing applications. Code which is based on N-Ns is generally much easier to write and maintain. Additionally, messages used to exchange data owned by other blocks may not have to travel long distances within the network, provided that the blocks are mapped physically close together in hardware. Communication, even on modern computing systems, is far more expensive than computation. Therefore, an initial strategy for domain decomposition is to enforce N-N dependencies between the blocks. In order to decompose a problem into smaller, independent pieces, we separate the global domain into blocks that share borders with their nearest neighbors. For example, in the case of a 1-D problem defined on the interval $[a,b]$ we can form $N$ blocks by writing
\begin{equation*}
    a = c_{0} < c_{1} < c_{2} < \cdots < c_{N} = b,
\end{equation*}
with $\Delta c_{i} = c_{i+1} - c_{i}$ denoting the width of block $i$. Multidimensional problems can be addressed in a similar way by partitioning the domain along multiple directions. Solving a PDE on each of these blocks, independently, requires an understanding of the various data dependencies. First, we address the local integrals $J_{*}$. Depending on the quadrature method, the reconstruction algorithm might require data from neighboring blocks. Reconstructions based on previously described WENO-type quadratures require an extension of the grid in order to build the interpolant. This involves a ``halo" region (see \cref{fig:quad_DD}), which is distributed amongst N-N blocks in the decomposition. On the other hand, more compact quadratures, such as Simpson's method \cite{causley2013method}, do not require this data. In this case, the quadrature communication phase can be ignored.

The major task for this work involves efficiently communicating the data necessary to build each of the convolution integrals $I_{*}$. Once the local integrals $J_{*}$ are constructed through quadrature, we sweep across the lines of the domain to build the convolution integrals. It is this operation which couples the integrals across the blocks. To decompose this operation, we first rewrite the integral operators $I_{*}$, assuming we are in block $i$, as 
\begin{align*}
    I_{L}[v; \alpha](x) &= \alpha \int_{x}^{b} e^{-\alpha(s - x)} v(s) \,ds, \\
                        &= \alpha \int_{x}^{c_{i+1}} e^{-\alpha(s - x)} v(s) \,ds + \alpha \int_{c_{i+1}}^{c_{N}} e^{-\alpha(s - x)} v(s) \,ds,
\end{align*}
and
\begin{align*}
    I_{R}[v; \alpha](x) &= \alpha \int_{a}^{x} e^{-\alpha(x - s)} v(s) \,ds, \\
                        &= \alpha \int_{c_{0}}^{c_{i}} e^{-\alpha(x - s)} v(s) \,ds + \alpha \int_{c_{i}}^{x} e^{-\alpha(x - s)} v(s) \,ds. 
\end{align*}
These relations, which assume $x$ is within the interval $[c_{i}, c_{i+1}]$, elucidate the local and non-local contributions to the convolution integrals within block $i$. Using simple algebraic manipulations, we can expand the non-local contributions to find that

\noindent
\begin{minipage}{\linewidth}
\begin{align*}
    \int_{c_{i+1}}^{c_{N}} e^{-\alpha(s - x)} v(s) \,ds &= \sum_{j=i+1}^{N-1} \int_{c_{j}}^{c_{j+1}} e^{-\alpha(s - x)} v(s) \,ds, \\
    &= \sum_{j=i+1}^{N-1} e^{-\alpha(c_{j} - x)} \int_{c_{j}}^{c_{j+1}} e^{-\alpha(s - c_{j})} v(s) \,ds,
\end{align*}
\end{minipage}
and
\begin{align*}
    \int_{c_{0}}^{c_{i}} e^{-\alpha(x - s)} v(s) \,ds &= \sum_{j=0}^{i-1} \int_{c_{j}}^{c_{j+1}} e^{-\alpha(x - s)} v(s) \,ds, \\
    &= \sum_{j=0}^{i-1} e^{-\alpha(x - c_{j+1})} \int_{c_{j}}^{c_{j+1}} e^{-\alpha(c_{j+1} - s)} v(s) \,ds,
\end{align*}
for the right and left-moving data, respectively. With these relations, each of the convolution integrals can be formed according to
\begin{equation}
    \label{eq:Left-moving integral DD}
    I_{L}[v; \alpha](x) = \alpha \int_{x}^{c_{i+1}} e^{-\alpha(s - x)} v(s) \,ds + \alpha \left( \sum_{j=i+1}^{N-1} e^{-\alpha(c_{j} - x)} \int_{c_{j}}^{c_{j+1}} e^{-\alpha(s - c_{j})} v(s) \,ds \right),
\end{equation}
and
\begin{equation}
    \label{eq:Right-moving integral DD}
    I_{R}[v; \alpha](x) = \alpha \left( \sum_{j=0}^{i-1} e^{-\alpha(x - c_{j+1})} \int_{c_{j}}^{c_{j+1}} e^{-\alpha(c_{j+1} - s)} v(s) \,ds \right) + \alpha \int_{c_{i}}^{x} e^{-\alpha(x - s)} v(s) \,ds.
\end{equation}
From equations \eqref{eq:Left-moving integral DD} and \eqref{eq:Right-moving integral DD}, we observe that both of the global convolution integrals can be split into a localized convolution with additional contributions coming from preceding or successive \textit{global integrals} owned by other blocks in the decomposition. These global integrals contain exponential attenuation factors, the size of which depends on the respective distances between any pair of sub-domains. Next, we use this result to derive the restriction that facilitates N-N dependencies.

\subsection{Nearest-Neighbor Criterion}
\label{subsec: NN criterion}

Building a consistent block-decomposition for the convolution integral is non-trivial, since this operation globally couples unknowns along a \textit{dimension} of the grid. Fortunately, the exponential kernel used in these reconstructions is pleasant in the sense that it automatically generates a region of compact support around a given block. Examining the exponential attenuation factors in \eqref{eq:Left-moving integral DD} and \eqref{eq:Right-moving integral DD}, we see that contributions from blocks beyond N-Ns become small provided that (1) the distance between the blocks is large or (2) $\alpha$ is taken to be sufficiently large. Since we have less control over the block sizes e.g., $\Delta c_{i}$, we can enforce the latter criterion. That is, we constrain $\alpha$ so that
\begin{equation}
    \label{eq:NN constraint general}
    e^{-\alpha L_{m}} \leq \epsilon,
\end{equation}
where $\epsilon \ll 1$ is some prescribed error tolerance, typically taken as $1\times10^{-16}$, and $L_{m} = \min_{i} \Delta c_{i}$ denotes the length smallest block. Taking logarithms of both sides and rearranging the inequality, we obtain the bound
\begin{equation*}
    - \alpha \leq \frac{\log(\epsilon)}{L_{m}}.
\end{equation*}

Our next step is to write this in terms of the time step $\Delta t$, using the choice of $\alpha$. However, the bound on the time step depends on the choice of $\alpha$. In \cref{subsec:Connections}, we presented two definitions for the parameter $\alpha$, namely
\begin{equation*}
    \alpha \equiv \frac{\beta}{c_{\text{max}} \Delta t}, \text{ or } \alpha \equiv \frac{\beta}{\sqrt{\nu \Delta t}}. 
\end{equation*}
Using these definitions for $\alpha$, we obtain two conditions depending on the choice of $\alpha$. For the linear advection equation and the wave equation, we obtain the condition
\begin{equation}
    \label{eq:NN-DD constraint 1}
     -\frac{\beta}{c_{\text{max}} \Delta t} \leq \frac{\log(\epsilon)}{L_{m}} \implies \Delta t \leq -\frac{\beta L_{m}}{ c_{\text{max}} \log(\epsilon)}.
\end{equation}
Likewise, for the diffusion equation, the restriction is given by
\begin{equation}
    \label{eq:NN-DD constraint 2}
     -\frac{\beta}{\sqrt{\nu \Delta t}} \leq \frac{\log(\epsilon)}{L_{m}} \implies \Delta t \leq \frac{1}{\nu} \left( \frac{\beta L_{m}}{\log(\epsilon)} \right)^{2}.
\end{equation}
Depending on the problem, if the condition \eqref{eq:NN-DD constraint 1} or \eqref{eq:NN-DD constraint 2} is not satisfied, then we use the maximally allowable time step for a given tolerance $\epsilon$, which is given by the equality component of the relevant condition.
% If $\Delta t$ is too large and the corresponding condition is not met, then we use the maximum allowable time step, i.e,  
% \begin{equation*}
%     \Delta t = -\frac{\beta L_{m}}{ c_{\text{max}} \log(\epsilon)}, \text{ or } \Delta t = \frac{1}{\nu} \left( \frac{\beta L_{m}}{\log(\epsilon)} \right)^{2}.
% \end{equation*}
% or 
% \begin{equation*}
%     \Delta t = \frac{1}{\nu} \left( \frac{\beta L_{m}}{\log(\epsilon)} \right)^{2}.
% \end{equation*}
If several different operators appear in a given problem and are to be approximated with successive convolution, then each operator will be associated with its own $\alpha$. In such a case, we should bound the time step according to the condition that is more restrictive among \eqref{eq:NN-DD constraint 1} and \eqref{eq:NN-DD constraint 2}, which can be accomplished through the choice
\begin{equation}
    \label{eq:NN-DD composite constraint}
    \Delta t \leq \min \left( -\frac{\beta L_{m}}{ c_{\text{max}} \log(\epsilon)}, \frac{1}{\nu} \left( \frac{\beta L_{m}}{\log(\epsilon)} \right)^{2} \right).
\end{equation}
As before, when the condition is not met, then we use the equality in \eqref{eq:NN-DD composite constraint}.

Restricting $\Delta t$ according to \eqref{eq:NN-DD constraint 1}, \eqref{eq:NN-DD constraint 2}, or \eqref{eq:NN-DD composite constraint} ensures that contributions to the right and left-moving convolution integrals, beyond N-Ns, become negligible. This is important because it significantly reduces the amount of communication, at the expense of a potentially restrictive time step. Note that in \cref{subsec:CFL Studies}, we analyze the limitations of such restrictions for the linear advection equation. In our future work, we shall consider generalizations of our approach, which do not require \eqref{eq:NN-DD constraint 1}, \eqref{eq:NN-DD constraint 2}, or \eqref{eq:NN-DD composite constraint}. In \cref{subsec: NN DD for partial x,subsec: NN DD for partial xx}, we demonstrate how to formulate block-wise definitions of the global $\mathcal{L}_{*}^{-1}$ operators using the derived conditions \eqref{eq:NN-DD constraint 1}, \eqref{eq:NN-DD constraint 2}, or \eqref{eq:NN-DD composite constraint}. 

\subsection{Enforcing Boundary Conditions for \texorpdfstring{$\partial_{x}$}{}}
\label{subsec: NN DD for partial x}

\begin{figure}[t]
    \centering
    \includegraphics[width=0.65\linewidth]{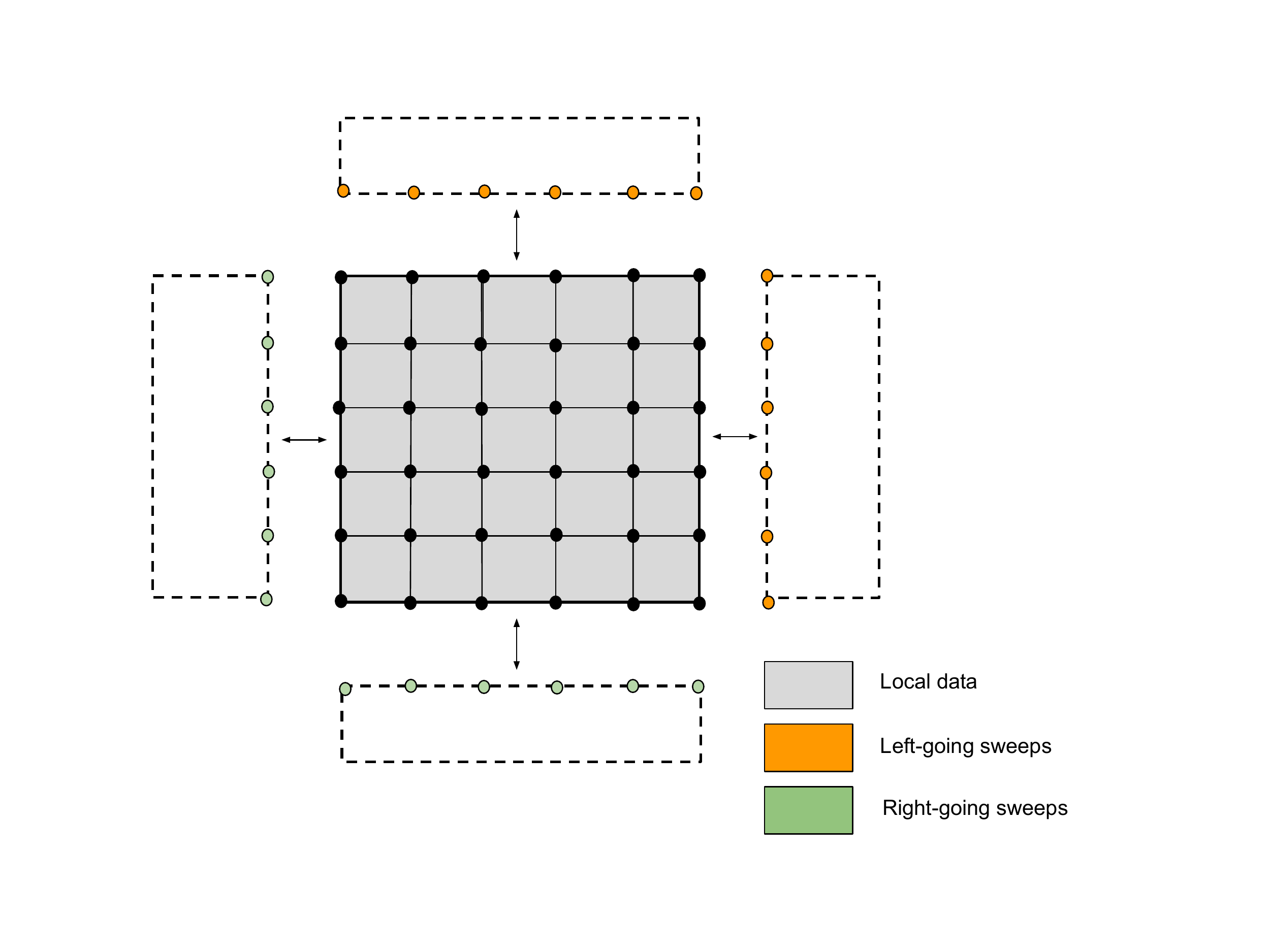}
    \caption{Fast convolution communication stencil in 2D based on N-Ns.}
    \label{fig:FC_DD}
\end{figure}

In order to enforce the block-wise boundary conditions for the first derivative $\partial_{x}$, we recall our definitions \eqref{eq:left-moving inverse op first derivative} and \eqref{eq:right-moving inverse op first derivative} for the left and right-moving inverse operators:
\begin{align}
    \label{eq:DD left going global true inverse op}
    \mathcal{L}_{L}^{-1}[v;\alpha](x) &= I_{L}[v;\alpha](x) + Be^{-\alpha (b - x)}, \\
    \label{eq:DD right going global true inverse op}
    \mathcal{L}_{R}^{-1}[v;\alpha](x) &= I_{R}[v;\alpha](x) + Ae^{-\alpha (x - a)}. 
\end{align}
We can modify these definitions so that each block contains a pair of inverse operators given by
\begin{align}
    \label{eq:Right inverse op DD}
    \mathcal{L}_{L,i}^{-1}[v;\alpha](x) &= I_{L,i}[v;\alpha](x) + B_{i}e^{-\alpha (c_{i+1} - x)}, \\
    \label{eq:Left inverse op DD}
    \mathcal{L}_{R,i}^{-1}[v;\alpha](x) &= I_{R,i}[v;\alpha](x) + A_{i}e^{-\alpha (x - c_{i})},
\end{align}
where $I_{*,i}$ are defined as
\begin{equation}
    \label{eq:local convolution integral defs}
    I_{L,i}[v;\alpha](x) = \alpha \int_{x}^{c_{i+1}} e^{-\alpha (s - x)} v(s) \, ds, \quad I_{R,i}[v;\alpha](x) = \alpha \int_{c_{i}}^{x} e^{-\alpha (x - s)} v(s) \, ds, 
\end{equation}
and the subscript $i$ denotes the block in which the operator is defined. As before, this assumes $x \in [c_{i},c_{i+1}]$. To address the boundary conditions, we need to determine expressions for the constants $B_{i}$ and $A_{i}$ on each of the blocks in the domain. First, substitute the definitions \eqref{eq:Left-moving integral DD} and \eqref{eq:Right-moving integral DD} into \eqref{eq:DD left going global true inverse op} and \eqref{eq:DD right going global true inverse op}:
\begin{align}
    \mathcal{L}_{L}^{-1}[v;\alpha](x) &= \alpha \int_{x}^{c_{i+1}} e^{-\alpha(s - x)} v(s) \,ds  \label{eq:integral def in left-moving inverse} \\
    &\hspace{2em} + \alpha \left( \sum_{j=i+1}^{N-1} e^{-\alpha(c_{j} - x)} \int_{c_{j}}^{c_{j+1}} e^{-\alpha(s - c_{j})} v(s) \,ds \right) + B e^{-\alpha (c_{N} - x)}, \nonumber \\
    \mathcal{L}_{R}^{-1}[v;\alpha](x) &= \alpha \left( \sum_{j=0}^{i-1} e^{-\alpha(x - c_{j+1})} \int_{c_{j}}^{c_{j+1}} e^{-\alpha(c_{j+1} - s)} v(s) \,ds \right) \label{eq:integral def in right-moving inverse} \\
    &\hspace{2em} + \alpha \int_{c_{i}}^{x} e^{-\alpha(x - s)} v(s) \,ds + A e^{-\alpha (x - c_{0})}. \nonumber
\end{align}
Since we wish to maintain consistency with the true operator being inverted, we require that each of the block-wise operators satisfy
\begin{equation*}
    \mathcal{L}_{L}^{-1}[v;\alpha](x) = \mathcal{L}_{L,i}^{-1}[v;\alpha](x), \quad \mathcal{L}_{R}^{-1}[v;\alpha](x) = \mathcal{L}_{R,i}^{-1}[v;\alpha](x),
\end{equation*}
which can be explicitly written as
\begin{align}
    \label{eq:DD left op consistency}
    B_{i} e^{-\alpha(c_{i+1} - x)} &= \left( \sum_{j=i+1}^{N-1} e^{-\alpha(c_{j} - x)} I_{L,j}[v;\alpha](c_{j}) \right) + B e^{-\alpha (c_{N} - x)}, \\
    \label{eq:DD right op consistency}
    A_{i} e^{-\alpha(x - c_{i})} &= \left( \sum_{j=0}^{i-1} e^{-\alpha(x - c_{j+1})} I_{R,j}[v;\alpha](c_{j+1}) \right) + A e^{-\alpha (x - c_{0})}.
\end{align}
Evaluating \eqref{eq:DD left op consistency} at $c_{i+1}$ and \eqref{eq:DD right op consistency} at $c_{i}$, we obtain
\begin{align}
    \label{eq:NN DD left op consistency}
    B_{i} &= \left( \sum_{j=i+1}^{N-1} e^{-\alpha(c_{j} - c_{i+1})} I_{L,j}[v;\alpha](c_{j}) \right) + B e^{-\alpha (c_{N} - c_{i+1})}, \\
    \label{eq:NN DD right op consistency}
    A_{i} &= \left( \sum_{j=0}^{i-1} e^{-\alpha(c_{i} - c_{j+1})} I_{R,j}[v;\alpha](c_{j+1}) \right) + A e^{-\alpha (c_{i} - c_{0})}.
\end{align}
Modifying $\Delta t$ according to either \eqref{eq:NN-DD constraint 1} or, if necessary \eqref{eq:NN-DD composite constraint}, results in the communication stencil shown in \cref{fig:FC_DD}. More specifically, the terms representing the boundary contributions in each of the blocks are given by
\begin{equation*}
    B_{i} =
    \begin{cases}
        B, \quad i = N - 1, \\[10pt]
        I_{R,i+1}[v;\alpha](c_{i+1}), \quad i < N - 1,
    \end{cases}
\end{equation*}
and
\begin{equation*}
    A_{i} =
    \begin{cases}
        A, \quad i = 0, \\[10pt]
        I_{R,i-1}[v;\alpha](c_{i}), \quad 0 < i. 
    \end{cases}
\end{equation*}
These relations generalize the various boundary conditions set by a problem. For example, with periodic problems, we can select
\begin{equation*}
    B = I_{L,0}[v;\alpha](c_{0}), \quad A = I_{R,N-1}[v;\alpha](c_{N}).
\end{equation*}
This is the relevant strategy employed by domain decomposition algorithms in this work. 

\subsection{Enforcing Boundary Conditions for \texorpdfstring{$\partial_{xx}$}{}}
\label{subsec: NN DD for partial xx}

The enforcement of boundary conditions on blocks of the domain for the second derivative can be accomplished using an identical procedure to the one described in \cref{subsec: NN DD for partial x}. First, we recall the inverse operator associated with a second derivative \eqref{eq:second derivative inverse op}:
\begin{equation}
    \label{eq:DD partial xx global true inverse op}
    \mathcal{L}_{0}^{-1}[v;\alpha](x) = I_{0}[v;\alpha](x) + A e^{-\alpha (x - a)} + B e^{-\alpha (b - x)},
\end{equation}
and define an analogous block-wise definition of \eqref{eq:DD partial xx global true inverse op} as
\begin{equation*}
    \mathcal{L}_{0,i}^{-1}[v;\alpha](x) = I_{0,i}[v;\alpha](x) + A_{i} e^{-\alpha (x - c_{0})} + B_{i} e^{-\alpha (c_{N} - x)},
\end{equation*}
with the localized convolution integral
\begin{equation*}
    I_{0,i}[v;\alpha](x) = \frac{\alpha}{2} \int_{c_{i}}^{c_{i+1}} e^{-\alpha \lvert x - s \rvert } v(s) \, ds.
\end{equation*}
Again, the subscript $i$ denotes the block in which the operator is defined and we take $x \in [c_{i},c_{i+1}]$. For the purposes of the fast summation algorithm, it is convenient to split this integral term into an average of left and right contributions, i.e.,
\begin{equation}
    \label{eq:DD partial xx local inverse op}
    \mathcal{L}_{0,i}^{-1}[v;\alpha](x) = \frac{1}{2} \Bigg( I_{L,i}[v;\alpha](x) + I_{R,i}[v;\alpha](x) \Bigg) + A_{i}e^{-\alpha (x - c_{i})} + B_{i}e^{-\alpha (c_{i+1} - x)},
\end{equation}
where $I_{*,i}$ are the same integral operators shown in equation \eqref{eq:local convolution integral defs} used to build the first derivative. As in the case of the first derivative, a condition connecting the boundary conditions on the blocks to the non-local integrals can be derived, which, if evaluated at the ends of the block, results in the $2 \times 2$ linear system
\begin{align}
    \label{eq:NN DD partial xx consistency eqn 1}
    A_{i} + B_{i}e^{-\alpha \Delta c_{i} } &= \frac{1}{2} \sum_{j=i+1}^{N-1} e^{-\alpha(c_{j} - c_{i})} I_{L,j}[v;\alpha](c_{j})  \\
    &+ \frac{1}{2} \sum_{j=0}^{i-1} e^{-\alpha(c_{i} - c_{j+1})} I_{R,j}[v;\alpha](c_{j+1}) \nonumber \\
    &+ A e^{-\alpha (c_{i} - c_{0})} + B e^{-\alpha (c_{N} - c_{i})}, \nonumber \\
    \label{eq:NN DD partial xx consistency eqn 2}
    A_{i}e^{-\alpha \Delta c_{i} } + B_{i} &= \frac{1}{2} \sum_{j=i+1}^{N-1} e^{-\alpha(c_{j} - c_{i+1})} I_{L,j}[v;\alpha](c_{j}) \\
    &+ \frac{1}{2} \sum_{j=0}^{i-1} e^{-\alpha(c_{i+1} - c_{j+1})} I_{R,j}[v;\alpha](c_{j+1})  \nonumber \\ 
    &+ A e^{-\alpha (c_{i+1} - c_{0})} + B e^{-\alpha (c_{N} - c_{i+1})}.  \nonumber
\end{align}
Equations \eqref{eq:NN DD partial xx consistency eqn 1} and \eqref{eq:NN DD partial xx consistency eqn 2} can be solved analytically to find that
\begin{equation*}
    \renewcommand{\arraystretch}{1.5}
    \begin{pmatrix}
        A_{i} \\
        B_{i}
    \end{pmatrix}
    = \frac{1}{1 - e^{-2\alpha \Delta c_{i}}}
    \begin{pmatrix}
        1 & -e^{-\alpha \Delta c_{i}} \\
        -e^{-\alpha \Delta c_{i}} & 1
    \end{pmatrix}
    \begin{pmatrix}
        r_{0} \\
        r_{1}
    \end{pmatrix},
\end{equation*}
where we have used the variables $r_{0}$ and $r_{1}$ to denote the terms appearing on the right-hand side of \eqref{eq:NN DD partial xx consistency eqn 1} and \eqref{eq:NN DD partial xx consistency eqn 2}, respectively. Under the N-N constraints \eqref{eq:NN-DD constraint 2} or \eqref{eq:NN-DD composite constraint}, many of the exponential terms can be neglected resulting in the compact expressions
\begin{equation*}
    A_{i} =
    \begin{cases}
         A, \quad i = 0, \\[10pt]
         \frac{1}{2} I_{R,i-1}[v;\alpha](c_{i}) \equiv I_{0,i-1}[v;\alpha](c_{i}), \quad 0 < i,
    \end{cases}
\end{equation*}
and
\begin{equation*}
    B_{i} =
    \begin{cases}
        B, \quad i = N - 1, \\[10pt]
        \frac{1}{2}I_{L,i+1}[v;\alpha](c_{i+1}) \equiv I_{0,i+1}[v;\alpha](c_{i+1}), \quad i < N - 1.
    \end{cases}
\end{equation*}

\subsection{Additional Comments}
\label{subsec: Sec 3 Additional Comments}

In this section, we developed the mathematical framework behind our proposed domain decomposition algorithm. We derived a condition, which reduces the construction of a nonlocal operator to a N-N dependency by leveraging the decay properties of the exponential term within the convolution integrals. We wish to reiterate that this condition is not entirely necessary. One could remove this condition by including contributions beyond N-Ns at the expense of additional communication. This change would certainly result in a loss of speed per time step, but the additional expense could be amortized by the ability to use much a larger time step, which would reduce the overall time-to-solution. As a first pass, we shall ignore these additional contributions, which may limit the scope of problems we can study, but we plan to generalize these algorithms in our future work via an adaptive strategy. This approach would begin using data from N-Ns, then gradually include additional contributions using information about the decay from the exponential. In the next section, we shall discuss details regarding the implementation of our methods and particular design choices made in the construction of our algorithms.

%
% Section 4
%

\section{Strategies for Efficient Implementation on Parallel Systems}
\label{sec:Implementation}

In this section, we discuss strategies for constructing parallel algorithms to solve PDEs. We provide the details related to our work on \textit{thread-scalable}, shared memory algorithms, as well as distributed memory algorithms, where the problem is decomposed into smaller, independent problems that communicate necessary information via message-passing. \Cref{subsec:Kokkos} introduces the core concepts in Kokkos performance portability library, which is used to develop our shared memory algorithms. Once we have introduced these ideas, we explore numerous loop-level optimizations for essential loop structures in \cref{subsec: Loop Optimizations} using the performance metrics discussed in \cref{subsec:Performance metrics}. Building on the results of these loop experiments, we outline the structure of our shared memory algorithms in \cref{subsec: Shared Memory Algorithms}. We then discuss the implementation of the distributed memory component of our algorithms in \cref{subsec:Distributed implementation}, along with modifications which enable the use of an adaptive time stepping rule. Finally, we summarize the key findings and developments of the implementation which are used to conduct our numerical experiments.

% { \noindent \color{red} \textbf{Comment:} Not sure what to do with this... }

% In general, the latter becomes necessary for solving large problems, which may consume more memory than what is available on a single node. Aside from memory, another reason we might consider developing distributed algorithms is because they distribute work across additional compute nodes in a supercomputer. The former programming model targets the parallelism available \textit{within} a given node, where the threads are allowed to execute concurrently. Shared-memory models are often used in large parallel codes because they address some of the concerns related to load balancing and they avoid data duplication. The approach used in this work involves a combination of these techniques, a so-called \textit{hybrid} method.    

\subsection{Selecting a Shared Memory Programming Model}
\label{subsec:Kokkos}

Many programming models exist to address the aspects of shared memory paralellization, such as OpenMP, OpenACC, CUDA, and OpenCL. The question of which model to use often depends on the target architecture on which the code is to run. However, given the recent trend towards deploying more heterogeneous computing systems, e.g., ones in which a given node contains a variety of CPUs with one, or many, accelerators (typically GPUs), the choice becomes far more complicated. Developing codes which are performant across many computing architectures is a highly non-trivial task. Due to memory access patterns, code which is optimized to run on CPUs is often not optimal on GPUs, so these models address \textit{portability} rather than \textit{performance}. This introduces yet another concern related to code management and maintenance: As new architectures are deployed, code needs to be tuned or modified to take advantage of new features, which can be time consuming. Additionally, enabling these abstractions almost invariably results in either multiple versions of the code or rather complicated build systems.

\begin{figure}[t]
    \centering
    \includegraphics[width=0.55\linewidth]{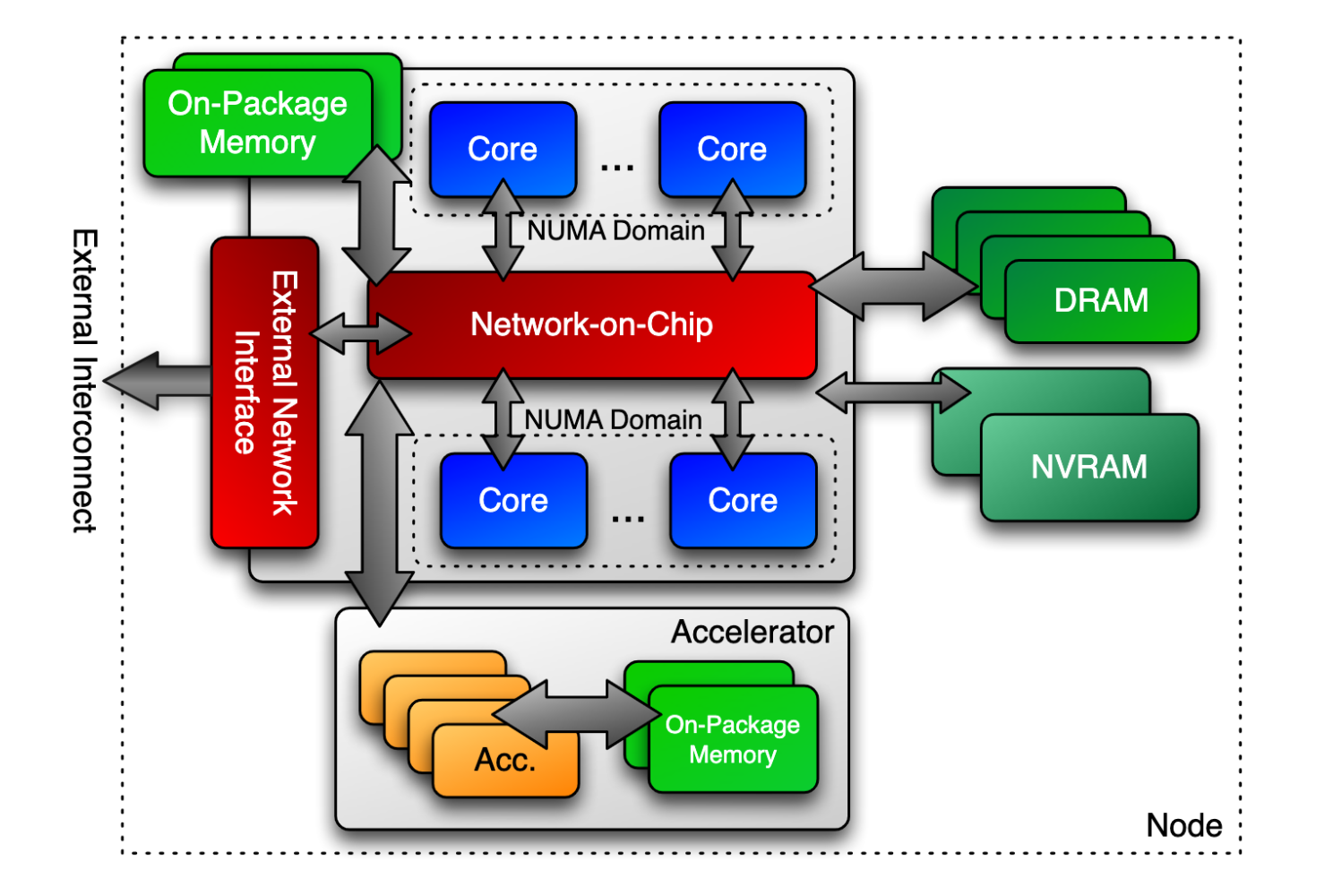}
    \caption{Heterogeneous platform targeted by Kokkos \cite{kokkos2014}} 
    \label{fig:target_arch}
\end{figure}

In our work, we choose to adopt Kokkos \cite{kokkos2014}, a \textit{performance portable}, shared memory programming model. Kokkos tries to address the aforementioned problem posed by rapidly evolving architectures through template-metaprogramming abstractions. Their model provides abstractions for common parallel policies (i.e., for-loops, reductions, and scans), memory spaces, and execution spaces. The architecture specific details are hidden from the users through these abstractions, yet the setup allows the application programmer to take advantage of numerous performance-related features. Given basic knowledge in templates, operator overloads, as well as functors and lambdas, one can implement a variety of program designs. Also provided are the so-called views, which are powerful multi-dimensional array containers that allow Kokkos iterators to map data onto various architectures in a performant way. Additionally, the bodies of iterators become either a user-defined functor or a \texttt{KOKKOS_LAMBDA}, which is just a functor that is generated by the compiler. This allows users to maintain one version of the code which has the flexibility to run on various architectures, such as the one depicted in \cref{fig:target_arch}. Other performance portability models, such as RAJA \cite{RAJA}, work in a similar fashion as Kokkos, but they are less intrusive with regard to memory management. With RAJA, the user is responsible for implementing architecture dependent details such as array layouts and policies. In this sense, RAJA emphasizes portability, with the user being responsible for handling performance.

\subsection{Comment on Performance Metrics}
\label{subsec:Performance metrics}

In order to benchmark the performance of the algorithms, we need a descriptive metric that accounts for varying workloads among problem sizes. In our numerical simulations we use a time stepping rule so that problems with a smaller mesh spacing require more time steps, i.e., $\Delta t \sim \Delta x$. Therefore, one can either time the code for a fixed number of steps or track the number of steps in the entire simulation $t \in (0,T]$ and compute the average time per time step. We adopt the former approach throughout this work. To account for the varying workloads attributed to varying cells/grid points, we define the update rate as \mbox{Degrees-of-Freedom/node/s (DOF/node/s)}, which can be computed via
\begin{equation}
    \label{eq:performance metric def}
    \text{DOF/node/s} = \frac{\text{total variables} \times N^d}{\text{nodes} \times \left( \frac{\text{total time (s)}}{\text{total steps}} \right) },
\end{equation}
where $d$ is the number of spatial dimensions. This metric is a more general way of comparing the raw performance of the code, as it allows for simultaneous comparisons among linear or nonlinear problems with varying degrees of dimensionality and number of components. It also allows for a comparison, in terms of speed, against other classes of methods, such as finite element methods, where the workload on a given cell is allowed to vary according to the number of basis elements \footnote{Note that the update frequency does not account for error in the numerical solution. Certainly, in order to compare the efficiency of various methods, especially those that belong to different classes, one must take into account the quality of the solution. This would be reflected in, for example, an error versus time-to-solution plot.}. In \cref{subsec: Loop Optimizations}, we shall use this performance metric to benchmark a collection of techniques for prescribing parallelism across predominant loop structures in the algorithms for successive convolution.

\subsection{Benchmarking Prototypical Loop Patterns}
\label{subsec: Loop Optimizations}

Often, when designing shared memory algorithms, one has to make design decisions prescribing the way threads are dispatched to the available data. However, there are often many ways of accomplishing a given task. Kokkos provides a variety of parallel iteration techniques --- the selection of a particular pattern typically depends on the structure of the loop (perfectly or imperfectly nested) and the size of the loops. In \cite{grete2019kathena}, authors sought to optimize a recurring pattern, consisting of triple or quadruple nested for-loops, in the Athena\texttt{++} MHD code \cite{athena2016}. Their strategy was to use a flexible loop macro to test various loop structures across a range of architectures. Athena\texttt{++} was already optimized to run on Intel Xeon-Phi platforms, so they primarily focused on approaches for porting to GPUs which maintained this performance on CPUs. Our work differs in that we have not yet identified optimal loop patterns for CPUs or GPUs and algorithms used here contain at least two major prototypical loop patterns. At the moment, we are not focusing on optimizing for GPUs, but, we do our best to keep in mind possible performance-related issues associated with various parallelization techniques. Some examples of recurring loop structures, in successive convolution algorithms, for 3D problems, are provided in \hyperref[3D_pattern_1]{Listing~\ref*{3D_pattern_1}} and \hyperref[3D_pattern_2]{Listing~\ref*{3D_pattern_2}}. Technically, there are left and right-moving operators associated with each direction, but, for simplicity, we will ignore this in the pseudo-code.  

Another important note, we wish to make, concerns the storage of the operator data on the mesh. Since the operations are performed ``line-by-line" on potentially large multidimensional arrays, we choose to store the data in memory so that the sweeps are performed on the fastest changing loop variables. This allows us to avoid significant memory access penalties associated with reading and writing to arrays, as the entries of interest are now consecutive in memory \footnote{This is true when the memory space is that of the CPU (host memory). In device memory, these entries will be ``coalesced", which is the optimal layout for threading on GPUs. This mapping of indices, between memory spaces, is automatically handled by Kokkos.}. For example, suppose we have an $N-$dimensional array with indices $x_1, x_2, \cdots, x_N$, and we wish to construct an operator in the $x_1$ direction. Then, we would store this operator in memory as \texttt{operator(}$ x_2, \cdots, x_N, x_1$\texttt{)}. The loops appearing in \hyperref[3D_pattern_1]{Listing~\ref*{3D_pattern_1}} and \hyperref[3D_pattern_2]{Listing~\ref*{3D_pattern_2}} can then be permuted accordingly. Note that the solution variable \texttt{u(}$x_1, x_2, \cdots, x_N$\texttt{)} is not transposed and is a read-only quantity during the construction of the operators.

\begin{minipage}{0.9\linewidth}
\begin{lstlisting}[language = C++, caption = {Pattern used in the construction of local integrals, convolutions, and boundary steps.}, label=3D_pattern_1]
for(int ix = 0; ix < Nx; ix++){
  for(int iy = 0; iy < Ny; iy++){
    // Perform some intermediate calculations
    // ...
    // Apply 1D algorithm to z-line data 
    for(int iz = 0; iz < Nz; iz++){
        z_operator(ix,iy,iz) = ...
    }
  }
}
\end{lstlisting}
\end{minipage}

\begin{minipage}{0.9\linewidth}
\begin{lstlisting}[language = C++, caption = {Another pattern used to build ``resolvent" operators. With some modifications, this same pattern could be used for the integrator step. In several cases, this iteration pattern may require reading entries, which are separated by large distances (i.e., the data is strided), in memory.},
label=3D_pattern_2]
for(int ix = 0; ix < Nx; ix++){
  for(int iy = 0; iy < Ny; iy++){
    for(int iz = 0; iz < Nz; iz++){
      z_operator(ix,iy,iz) = u(ix,iy,iz) - z_operator(ix,iy,iz);
      z_operator(ix,iy,iz) *= alpha_z;
    }
  }
}
\end{lstlisting}
\end{minipage}

In an effort to develop an efficient application, we follow the approach described in \cite{grete2019kathena}, to determine optimal loop iteration techniques for patterns, such as \hyperref[3D_pattern_1]{Listing~\ref*{3D_pattern_1}} and \hyperref[3D_pattern_2]{Listing~\ref*{3D_pattern_2}}. Our simple 2D and 3D experiments tested numerous combinations of policies including naive, as well as more complex parallel iteration patterns using the OpenMP backend in Kokkos. Our goals were to quantify possible performance gains attainable through the following strategies: 
\begin{enumerate}[topsep=1em,itemsep=1em]
% \begin{enumerate}
    \item auto-vectorization via \texttt{\#pragma} statements or \texttt{ThreadVectorRange} (TVR)
    \item improving data reuse and caching behavior with loop tiling/blocking
    \item prescribing parallelism across combinations of team-type execution policies and team sizes
\end{enumerate}

\begin{figure}[!ht]

    \centering
    \subfloat{\label{fig:2D_J_loop_exp}}{\includegraphics[width=0.42\linewidth]{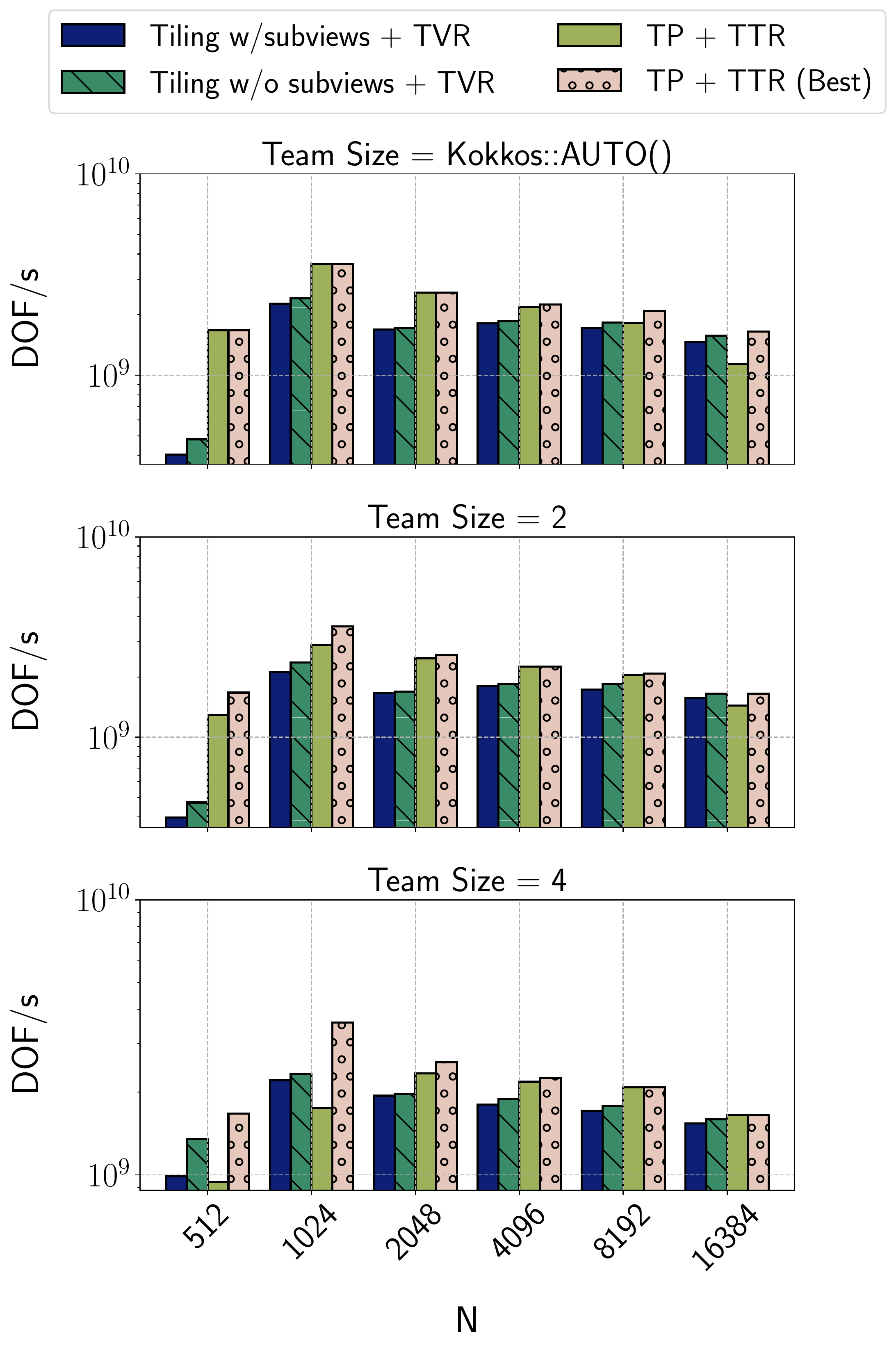}}
    \hfill
    \subfloat{\label{fig:3D_J_loop_exp}}{\includegraphics[width=0.56\linewidth]{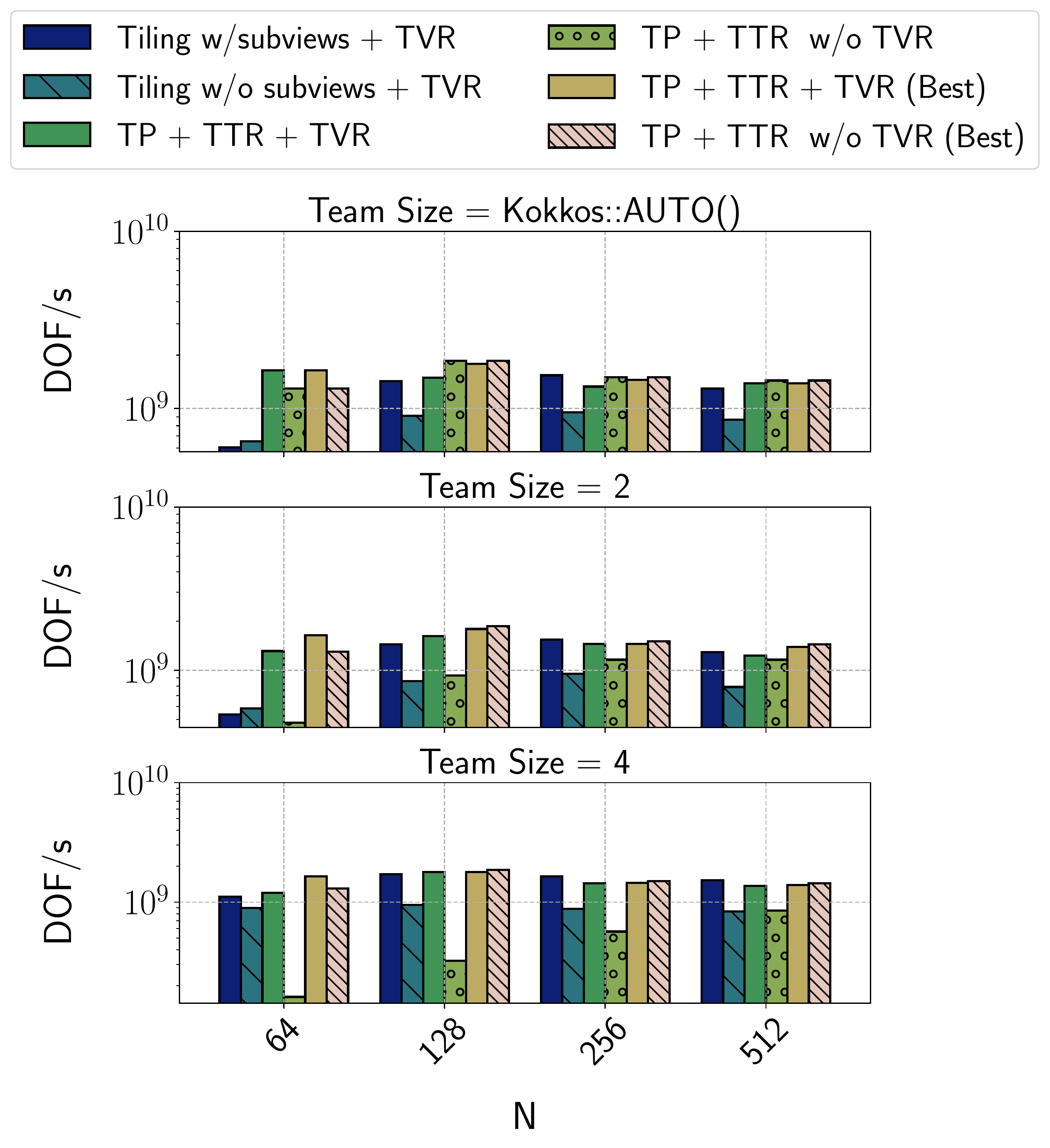}}
    \caption{Plots comparing the performance of different parallel execution policies for the pattern in \hyperref[3D_pattern_1]{Listing~\ref*{3D_pattern_1}} using test cases in 2D (left) and 3D (right). Tests were conducted on a single node that consists of 40 cores using the code configuration outlined in \cref{tab:compiler and opt flags}. Each group consists of three plots, whose difference is the value selected for the team size. We note that hyperthreading is not enabled on our systems, so \texttt{Kokkos::AUTO()} defaults to a team size of 1. In each pane, we use ``best" to refer to the best run for that configuration across different team sizes. Tile experiments used block sizes of $256^2$, in 2D problems, and $32^3$ in 3D. We observe that vectorized policies are generally faster than non-vectorized policies. Interestingly, among blocked/tiled policies, construction of subviews appears to be faster than those that skip the subview construction, despite the additional work. As the problem size increases, the performance of blocked policies improves substantially. This can be attributed to the large number of idle thread teams when the problem size does not produce enough blocks. In such cases, increasing the size of the team does offer an improvement, as it reduces the number of idle thread teams. For non-blocked policies, we observe that increasing the team-size generally results in minimal, if any, improvement in performance. In all cases, the use of blocking provides a more consistent update rate when enough work is introduced.}
    \label{fig:J_loop_exp}
\end{figure}

\begin{figure}[!ht]
    
    \centering
    \subfloat{\label{fig:2D_D_loop_exp}}{\includegraphics[width=0.68\linewidth]{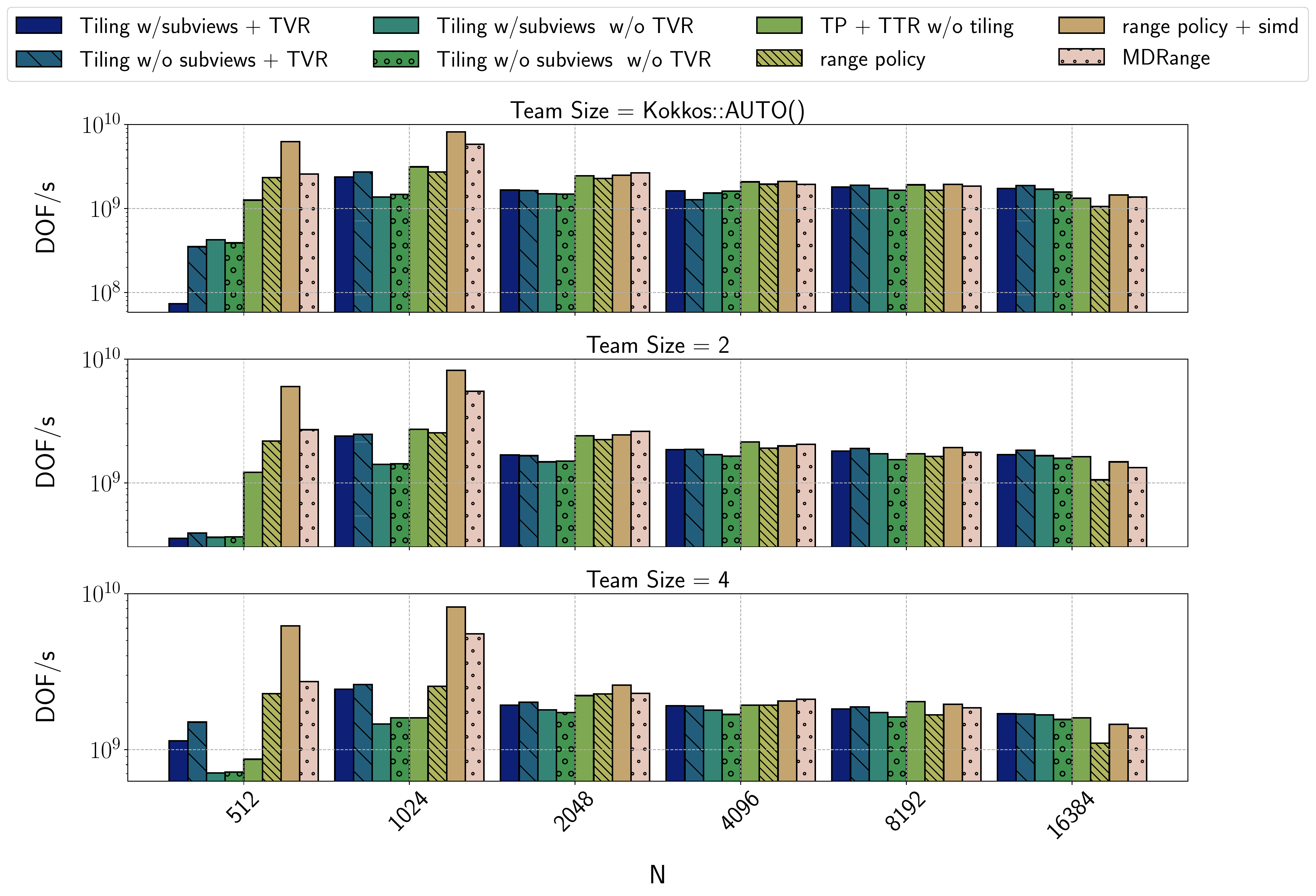}}
    \hfill
    \subfloat{\label{fig:3D_D_loop_exp}}{\includegraphics[width=0.63\linewidth]{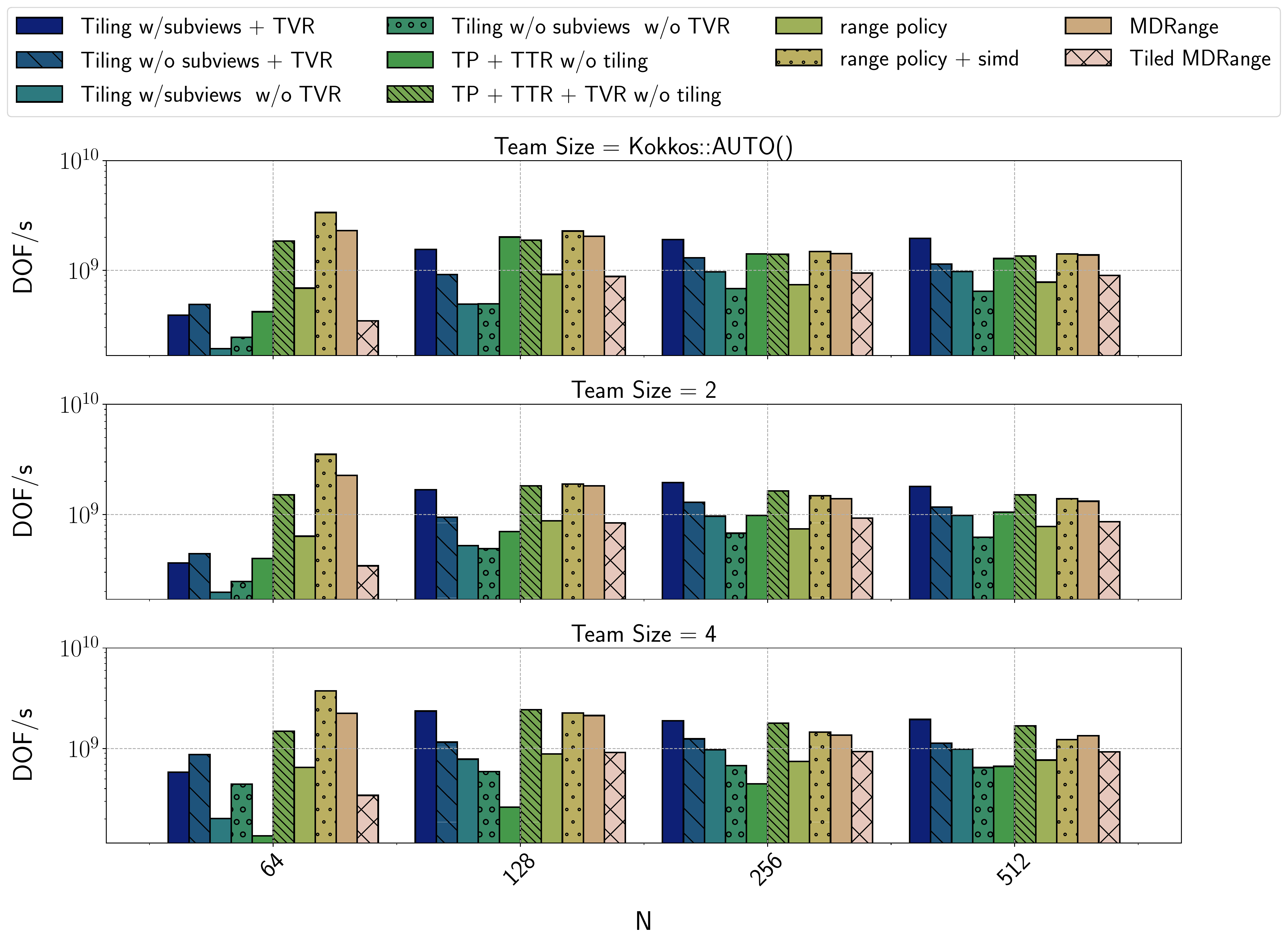}}
    \caption{Plots comparing the performance of different parallel execution policies for the pattern in \hyperref[3D_pattern_2]{Listing~\ref*{3D_pattern_2}} using test cases in 2D (top) and 3D (bottom). Tests were conducted on a single node that consists of 40 cores using the code configuration outlined in \cref{tab:compiler and opt flags}. Each group consists of three plots, whose difference is the value selected for the team size. We note that hyperthreading is not enabled on our systems, so \texttt{Kokkos::AUTO()} defaults to a team size of 1. Tile experiments used a block size of $256^2$, in 2D problems, and $32^3$ in 3D. A tiled MDRange was not implemented in the 2D cases because the block size was larger than some of the problems. The results generally agree with those presented in \cref{fig:J_loop_exp}. For smaller problem sizes, using the non-portable \texttt{range_policy} with OpenMP simd directives is clearly superior over the policies. However, when enough work is available, we see that blocked policies with subviews and vectorization generally become the fastest. In both cases, \texttt{MDRange} seems to have fairly good performance. Tiling, when used with \texttt{MDRange}, in the 3D cases, seems to be slower than plain \texttt{MDRange}. Again, we see that the use of blocking provides a more consistent update rate if enough work is available. }
    \label{fig:D_loop_exp}
\end{figure}

Vectorization can offer substantial performance improvements for data that is contiguous in memory. However, several performance critical operations in the algorithms involve reading data which is strided in memory. Therefore, it is not straightforward whether vectorization would offer any improvements. Additionally, for larger problems, the line operations along certain directions involve reading strided data, so that benefits of caching are lost. The performance penalty of operating on data with the wrong layout depends on the architecture, with penalties on GPUs typically being quite severe compared to CPUs. The use of a blocked iteration pattern, such as the one outlined in \hyperref[team_tiling]{Listing~\ref*{team_tiling}} (see \cref{sec:kokkos kernels}), is a step toward minimizing such performance penalties. In order to see a performance benefit from this approach, the algorithms must be structured, in such a way, as to reuse the data that is read into caches, as much as possible. Naturally, one could prescribe one or more threads (of a team) to process blocks, so we chose to implement cache blocking using the hierarchical execution policies provided by Kokkos. From coarse-to-fine levels of granularity, these can be ordered as follows: \texttt{TeamPolicy} (TP), \texttt{TeamThreadRange} (TTR), and \texttt{ThreadVectorRange} (TVR). For perfectly nested loops, one can achieve similar behavior using \texttt{MDRange} and prescribing block sizes. During testing, we found that when block sizes are larger than or equal to the size of the view, a segmentation fault occurs, so this was avoided. The results of our loop experiments are provided in \cref{fig:J_loop_exp,fig:D_loop_exp}. For tests employing blocking, we used a block size of $256^2$ in 2D, while 3D problems used a block size of $32^3$. Information regarding various choices, such as compiler, optimization flags, etc., used to generate these results can be found in \cref{tab:compiler and opt flags}.

\begin{table}[!ht]
\centering
\renewcommand{\arraystretch}{1.5}
\resizebox{\columnwidth}{!}{
\begin{tabular}{c || c} 
 \hline
 \textbf{CPU Type} & Intel Xeon Gold 6148 \\
 \textbf{C\texttt{++} Compiler} & ICC 2019.03 \\
 \textbf{Optimization Flags} & \texttt{-O3 -xCORE-AVX512 -qopt-zmm-usage=high -qno-opt-prefetch} \\
 \textbf{Thread Bindings} & \texttt{OMP_PROC_BIND=close, OMP_PLACES=threads} \\
 \hline 
\end{tabular}
}
\caption{Architecture and code configuration for the loop experiments conducted on the Intel 18 cluster at Michigan State University's Institute for Cyber-Enabled Research. To leverage the wide vector registers, we encourage the compiler to use AVX-512 instructions. Hardware prefetching is not used, as initial experiments seem to indicate that it hindered performance. Initially, we used GCC 8.2.0-2.31.1 as our compiler, but we found through experimentation that using an Intel compiler improved the performance of our application by a factor of $\sim 2$ for this platform. Authors in \cite{grete2019kathena} experienced similar behavior for their application and attribute this to a difference in auto-vectorization capabilities between compilers. An examination of the source code for loop execution policies in Kokkos reveals that certain decorators, e.g., \texttt{\#pragma ivdep} are present, which help encourage auto-vectorization when Intel compilers are used. We are unsure if similar hints are provided for GCC.}
\label{tab:compiler and opt flags}
\end{table}

As part of our blocking implementation, we stored block information in views, which could then be accessed by a team of threads. After the information about the block is obtained, we compute indices for the data within the block and use these to extract the relevant grid data. Then, one can either create subviews (shallow copies) of the block data or proceed directly with the line calculations of the block data. We refer to these as tiling with and without subviews, respectively. Intuitively, one would think that skipping the block subview creation step would be faster. Among the blocked or tiled experiments, those that created the subviews of the tile data were generally faster than those that did not. Using blocking for smaller problems typically resulted in a large number of idle threads, which significantly degraded the performance compared to non-blocked policies. In such situations, a user would need to take care to ensure that a sufficient number of blocks are used to generate enough work, i.e., each thread (or team) has at least one block to process. For larger problems, blocking was faster when compared to variants that did not use blocking. We observe that the performance of non-blocked policies begins to degrade once a problem becomes sufficiently large, whereas blocked policies maintained a consistent update rate, even as the problem size increased. By separating the key loop structures from the complexities of the application, we were able to expedite the experimental process for identifying efficient loop execution techniques. In \cref{subsec: Shared Memory Algorithms}, we use the results of these experiments to inform choices regarding the design of the shared memory algorithms.

% J loops

% D loops

\subsection{Shared Memory Algorithms}
\label{subsec: Shared Memory Algorithms}

The line-by-line approach to operator reconstruction suggests that we employ a \textit{hierarchical design}, which consists of thread teams. Rather than employ a fine-grained threading approach over loop indices, we use the coarse-grained, blocked iteration pattern devised in \cref{subsec: Loop Optimizations}. In this approach, we divide the iteration space into blocks of nearly identical size, and assign one or more blocks to a team of threads. The threads within a given team are then dispatched to one (or more) lines, with vector instructions being used within the lines. As opposed to loop level parallelism, coarse-grained approaches allow one to exploit multiple levels of parallelism, common to many modern CPUs, and load balance the computation across blocks by adjusting the loop scheduling policy. In our implementation, we provide the flexibility of setting the number of threads per block with a macro, but, in general, we let Kokkos choose the appropriate team size using \texttt{Kokkos::AUTO()}. If running on the CPU, this sets the team size to be the number of hyperthreads (if supported) on a given core. For GPU architectures, the team size is the size of the warp. 

A hierarchical design pattern is used because the loops in our algorithms are not perfectly nested, i.e., calculations are performed between adjoining loops. Information related to blocking can be precomputed to minimize the number of operations are required to manipulate blocks. The process of subview construction consists of shallow copies involving pointers to vertices of the blocks, so no additional memory is required. With a careful choice of a base block size, one can fit these blocks into high-bandwidth memory, so that accessing costs are reduced. Furthermore, a team-based, hierarchical pattern seems to provide a large degree of flexibility compared to standard loop-level parallelism. In particular, we can fuse adjacent kernels into a single parallel region, which reduces the effect of kernel launch overhead and minimizes the number of synchronization points. The use of a team-type execution policy also allows us to exploit features present on other architectures, such as CUDA's shared memory feature, through scratchpad constructs. Performing a stenciled operation on strided data is associated with an architecture-dependent penalty. On CPUs, while one wishes to operate in a contiguous or cached pattern, various compilers can hide these penalties through optimizations, such as prefetching. GPUs, on the other hand, prefer to operate in a lock-step fashion. Therefore, if a kernel is not vectorizable, then one pays a significant performance penalty for poor data access patterns. Shared memory, while slower than register accesses, does not require coalesced accesses, so the cost can be significantly reduced. The advantage of using square-like blocks of a fixed size, as opposed to long pencils \footnote{We refer to a pencil as a, generally, long rectangle (in 2-D) and a rectangular prism (in 3-D). The use of pencils, as opposed to square blocks, would require additional precomputing efforts and, possibly, restrictions on the problem size.}, is that one can adjust the dimensions of the blocks so that they fit into the constraints of the high-bandwidth memory. Moreover, these blocks can be loaded once and can be reused for additional directions, whereas pencils would require numerous transfers, as lines are processed along a given dimension. Such optimizations are not explored in this work, but algorithmic flexibility is something we must emphasize moving forwards.

The parallel nested loop structures, such as the one provided in \cref{sec:kokkos kernels} (see \hyperref[team_tiling]{Listing~\ref*{team_tiling}}) are applied during reconstructions for the local integrals $J_{*}$ and inverse operators $\mathcal{L}_{*}^{-1}$, as well as the integrator update. The current exception to this pattern is the convolution algorithm, shown in \hyperref[serial_FC]{Listing~\ref*{serial_FC}}, which is also provided in \cref{sec:kokkos kernels}. Here, each thread is responsible for constructing $I_{*}$ on one or more lines of the grid. Therefore, within a line, each thread performs the convolution sweeps, in serial, using our $\mathcal{O}(N)$ algorithm. Adopting the team-tiling approach for this operation requires that we modify our convolution algorithm considerably -- this optimization is left to future work. Additionally the benefit of this optimizations is not large for CPUs, as profiling indicated that $< 5\%$ of the total time for a given run was spent inside this kernel. However, this will likely consume more time on GPUs, so this will need to be investigated.  

If one wishes to use variable time stepping rules, where the time step is computed from a formula of the form
\begin{equation}
    \label{eq:Adaptive time rule}
    \Delta t = \text{CFL} \min\left(\frac{\Delta x}{c_{x}}, \frac{\Delta y}{c_{y}}, \cdots \right),
\end{equation}
then one must supply parallel loop structures with simultaneous maximum reductions for each of the wave speeds $c_{i}$. This can be implemented as a custom functor, but the use of blocking/tiling introduces some complexities. More complex reducers that enable such calculations are not currently available. For this reason, problems that use time stepping rules, such as \eqref{eq:Adaptive time rule}, are constructed with symmetry in the wave speeds, i.e., $c_x = c_y = \cdots$ to avoid an overly complex implementation with blocking/tiling. However, we plan to revisit this in later work as we begin targeting more general problems. Next, in \cref{subsec:Distributed implementation}, we discuss the distributed memory component of the implementation and the strategy used to employ an adaptive time stepping rule, such as \eqref{eq:Adaptive time rule}.

\subsection{Code Strategies for Domain Decomposition}

\label{subsec:Distributed implementation}

One of the issues with distributed computing involves mapping the problem data in an intelligent way so that it best aligns with the physical hardware. Since the kernels used in our algorithms consume a relatively small amount of time, it is crucial that we minimize the time spent communicating data. Given that these schemes were designed to run on Cartesian meshes, we can use a ``topology aware" virtual communicator supplied by MPI libraries. These constructs take a collection of ranks in a communicator (each of which manages a sub-domain) and, if permitted, attempt to reorganize them to best align with the physical hardware. This mapping might not be optimal, since it depends on a variety of factors related to the job allocation and the MPI implementation. Depending on the problem, these tools can greatly improve the performance of an application compared to a hand-coded implementation that uses the standard communicator. Additionally, MPI's Cartesian virtual communicator provides functionality to obtain neighbor references and derive additional communicator groups, say, along rows, columns, etc. The send and receive operations are performed with \textit{persistent communications}, which remove some of the overhead for communication channel creation. Persistent communications require a regular communication channel, which, for our purposes, is simply N-Ns. For more general problems with irregular communication patterns, standard send and receive operations can be used.

% For this case, we can use the following process flow:
\begin{figure}[t]
    \centering
    %trim=left bottom right top
    \subfloat{\label{fig:task_chart}}{\includegraphics[trim=120 10 120 10,clip, width=0.49\linewidth, keepaspectratio=true]{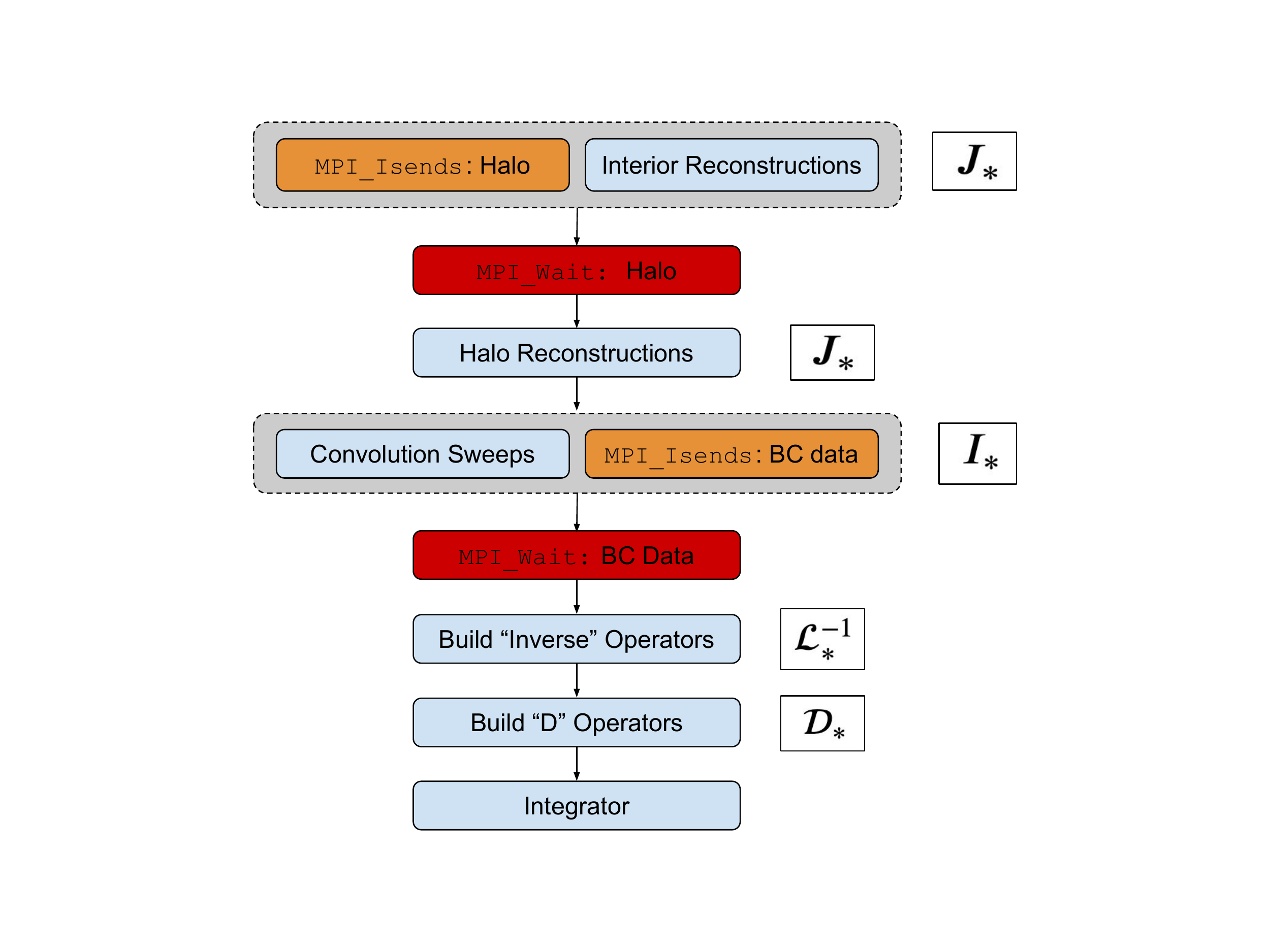}}
    \subfloat{\label{fig:asyc_task_chart}}{\includegraphics[trim=120 10 120 10,clip, width=0.49\linewidth,keepaspectratio=true]{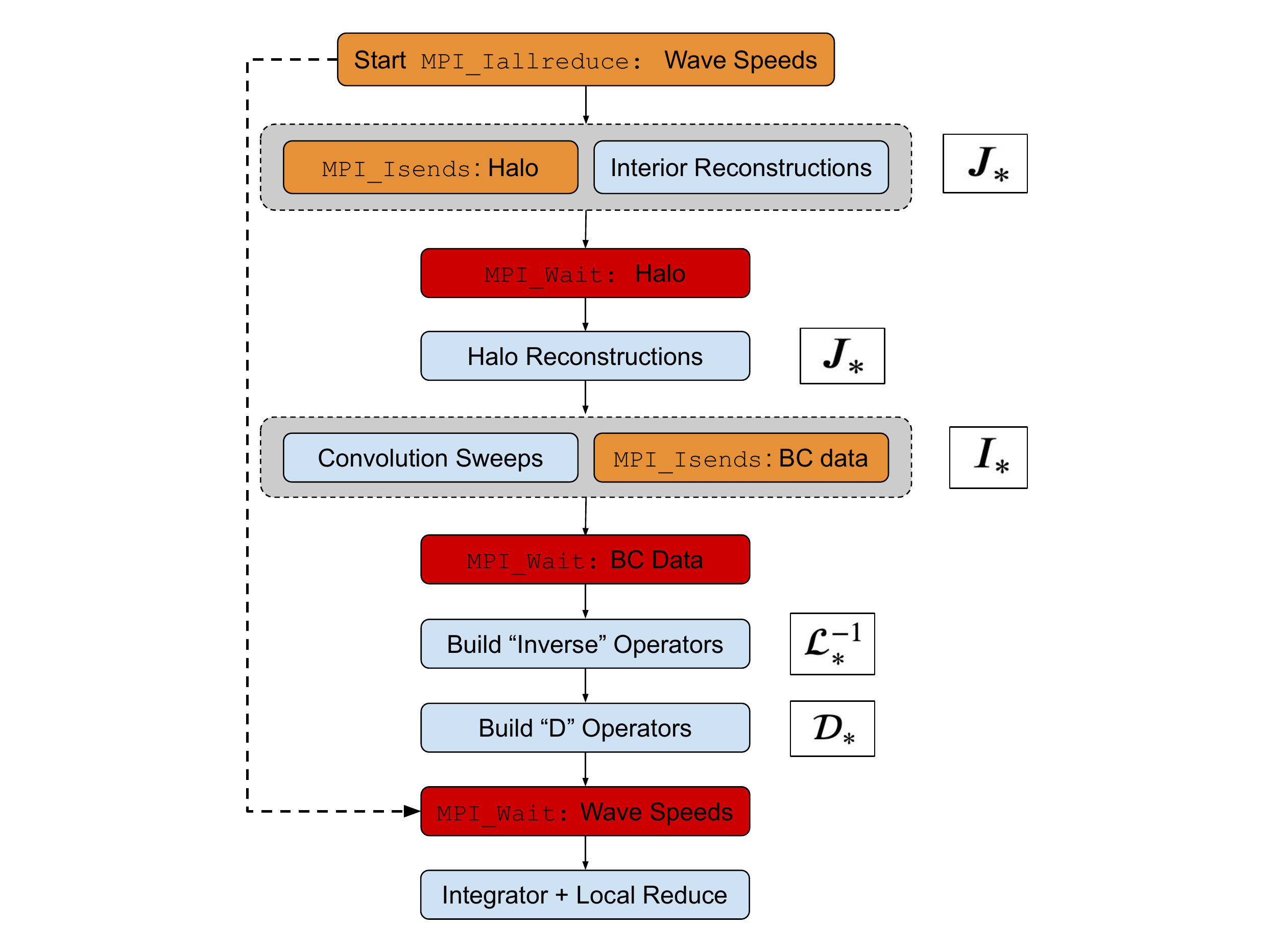}}
    \caption{Task charts for the domain-decomposition algorithm under fixed (left) and adaptive (right) time stepping rules. The work overlap regions are indicated, laterally, using gray boxes. The work inside the overlap regions should be sufficiently large to hide the communications occuring in the background. To clarify, the overlap in calculations for $I_{*}$ is achieved by changing the sweeping direction during an exchange of the boundary data. As indicated in the adaptive task chart, the reduction over the ``lagged" wave speed data can be performed in the background while building the various operators. Note the use of \texttt{MPI_WAIT} prior to performing the integrator step. This is done to prevent certain overwrite issues during the local reductions in the subsequent integrator step.}
    \label{fig:DD_task_charts}
\end{figure}

One strategy to minimize exposed communications is to use \textit{non-blocking} communications. This allows the programmer to overlap calculations with communications, and is especially beneficial if the application can be written in a staggered fashion. If certain data required for a later calculation is available, then communication can proceed while another calculation is being performed. Once the data is needed, we can block progress until the message transfer is complete. However, we hope that the calculation done, in between, is sufficient to hide the time spent performing the communication. In the multi-dimensional setting, other operators may be needed, such as $\partial_{x}$ and $\partial_{y}$. However, in our algorithms, directions are not coupled, which allows us to stagger the calculations. So, we can initialize the communications along a given direction and build pieces of other operators in the background. 

A typical complication that arises in distributed implementations of PDE solvers concerns the use of various expensive collective operations, such as ``all-to-one" and ``one-to-all" communications. For implicit methods, these operations occur as part of the iterative method used for solving distributed linear systems. The method employed here is ``matrix-free", which eliminates the need to solve such distributed linear systems. For explicit methods, these operations arise when an adaptive time stepping rule, such as equation \eqref{eq:Adaptive time rule}, is employed to ensure that the CFL restriction is satisfied for stability purposes. At each time step, each of the processors, or ranks, must know the maximum wave speeds across the entire simulation domain. On a distributed system, transferring this information requires the use of certain collective operations, which typically have an overall complexity of $\mathcal{O}(\log N_p)$, where $N_p$ is the number of processors. While the logarithmic complexity results in a massive reduction of the overall number of steps, these operations use a barrier, in which all progress stops until the operation is completed. This step cannot be avoided for explicit methods, as the most recent information from the solution is required to accurately compute the maximum wave speeds. In contrast, successive convolution methods, do not require this information. However, implementations for schemes developed in e.g., \cite{christlieb2019kernel, christlieb2020_NDAD, christlieb2020nonuniformHJE} considered ``explicit" time stepping rules given by equation \eqref{eq:Adaptive time rule} because they improved the convergence of the approximations. By exploiting the stability properties associated with successive convolution methods, we can eliminate the need for accurate wave speed information, based on the current state, and, instead, use approximations obtained with ``lagged" data from the previous time step. We present two, generic, distributed memory task charts in \cref{fig:DD_task_charts}. The algorithm shown in the left-half of \cref{fig:DD_task_charts}, which is based on a fixed time step, contains less overall communication, as the local and global reductions for certain information used to compute the time step are no longer necessary. The second version, shown in the right-half of \cref{fig:DD_task_charts}, illustrates the key steps used in the implementation of an adaptive rule, which can be used for problems with more dynamic quantities (e.g., wave speeds and diffusivity). In contrast to distributed implementations of explicit methods, our adaptive approach allows us to overlap expensive global collective operations (approximately) with the construction of derivative operators, resulting in a more asynchronous algorithm (see \cref{alg:adaptive rule}).

\begin{algorithm}[!ht]
    \caption{Distributed adaptive time stepping rule}
    Approximate the global maximum wave speeds $c_x, c_y, \cdots$ using the corresponding ``lagged" variables $\tilde{c}_x, \tilde{c}_y, \cdots$
    \label{alg:adaptive rule}
    \begin{algorithmic}[1]
    \State Initialize the $c_{i}$'s and $\tilde{c}_{i}$'s via the initial condition (no lag has been introduced) 
    \While{timestepping}
    
    \State \label{line3} Update the N-N condition (i.e., \eqref{eq:NN-DD constraint 1}, \eqref{eq:NN-DD constraint 2}, or \eqref{eq:NN-DD composite constraint}) using the ``lagged" wave speeds $\tilde{c}_x, \tilde{c}_y, \cdots$
    
    \State Compute $\Delta t$ using the ``lagged" wave speeds and check the N-N condition 
    
    \State Start the \texttt{MPI_Iallreduce} over the local wave speeds $c_x, c_y, \cdots$ 
    
    \State Construct the spatial derivative operators of interest
    
    \State Post the \texttt{MPI_WAIT} (in case the reductions have not completed)
    
    \State Transfer the global wave speed information to the corresponding lagged variables: 
        \begin{align*}
            \tilde{c}_x &\leftarrow c_x, \\
            \tilde{c}_y &\leftarrow c_y, \\
            &\vdots
        \end{align*}
    
    \State Perform the update step and computes the local wave speeds $c_x, c_y, \cdots$  
    
    \State Return to step \ref{line3} to begin the next time step 
    
    \EndWhile
    \end{algorithmic}
\end{algorithm}

% \begin{remark}

% The distributed adaptive time stepping rule, shown in Figure \ref{fig:asyc_task_chart}, approximates the global maximum wave speeds $c_x, c_y, \cdots$ using the corresponding ``lagged" variables $\tilde{c}_x, \tilde{c}_y, \cdots$. These values are first set according to the initial condition, at which point, no lag has been introduced. Within each time step, we proceed as follows:
% \begin{enumerate}[topsep=1em,itemsep=1em]
% %\begin{enumerate}
%     \item Estimate $\Delta t$ using the ``lagged" wave speeds $\tilde{c}_x, \tilde{c}_y, \cdots$ 
%     \item Start the non-blocking all reduce over the local wave speeds $c_x, c_y, \cdots$ and perform the constructions for derivative operators of interest.
%     \item Before entering the integrator step, post a block for the collective in case the operation has not completed.
%     \item To prevent overwrites, transfer the global wave speed information to the corresponding lagged variables
%     \begin{align*}
%         \tilde{c}_x &\leftarrow c_x, \\
%         \tilde{c}_y &\leftarrow c_y, \\
%         &\vdots
%     \end{align*}
%     \item Perform the integration step, which computes the local wave speeds $c_x, c_y, \cdots$ and updates the solution. 
%     \item Return to step (1) to begin the next time step.
% \end{enumerate}

% \end{remark}

\subsection{Some Remarks}

In this section, we introduced key aspects that are necessary in developing a performant application. We began with a brief discussion on Kokkos, which is the programming model used for our shared memory implementation. Then we introduced one of the metrics, namely \eqref{eq:performance metric def}, used to characterize the performance of our parallel algorithms. Using this performance metric, we analyzed a collection of techniques for parallelizing prototypical loop structures in our algorithms. These techniques considered several different approaches to the prescription of parallelism through both naive and complex execution policies. Informed by these results, we chose to adopt a coarse-grained, hierarchical approach that utilizes the extensive capabilities available on modern hardware. In consideration of our future work, this approach also offers a large degree of algorithmic flexibility, which will be essential for moving to GPUs. Finally, we provided some details concerning the implementation of the distributed memory components of the parallel algorithms. We introduced two different approaches: one based on a fixed time step, with minimal communication, and another, which exploits the stability properties of the representations and allows for adaptive time stepping rules. The next section provides numerical results, which demonstrate not only the performance and scalability of these algorithms, but also their versatility in addressing different PDEs.

\section{Numerical Results}
\label{sec:Results}

%Describe what is contained in this section.
This section provides the experimental results for our parallel algorithms using MPI and Kokkos, together, with the OpenMP backend. First, in \cref{subsec:Convergence}, we define several test problems and verify the rates of convergence for the hybrid algorithms described in \cref{sec:DD Algorithm,sec:Implementation}. Next, we provide both weak and strong scaling results obtained from each of the example problems discussed in \cref{subsec:Convergence}. \Cref{subsec:CFL Studies} provides some insight on issues faced by the distributed memory algorithms, in light of the N-N condition \eqref{eq:NN-DD constraint 1}, which was derived in \cref{subsec: NN criterion}. Unless otherwise stated, the results presented in this section were obtained using the configurations outlined in \cref{tab:numerical experiment config table}. Timing data presented in \cref{fig:large_weak_scaling,fig:small_weak_scaling,fig:small_strong_scaling} was collected using 10 trials for each configuration (problem size and node count) with the update metric \eqref{eq:performance metric def} being displayed relative to $10^{9}$ DOF/node/s. Each of these trials, evolved the numerical solution over 10 time steps. Error bars, collected from data involving averages, were computed using the sample standard deviation. 

\begin{table}[!ht]
\centering
\renewcommand{\arraystretch}{1.5}
\resizebox{\columnwidth}{!}{
\begin{tabular}{c || c} 
 \hline
 \textbf{CPU Type} & Intel Xeon Gold 6148 \\
 \textbf{C\texttt{++} Compiler} & ICC 2019.03 \\
 \textbf{MPI Library} & Intel MPI 2019.3.199 \\
 \textbf{Optimization Flags} & \texttt{-O3 -xCORE-AVX512 -qopt-zmm-usage=high -qno-opt-prefetch} \\
 \textbf{Thread Bindings} & \texttt{OMP_PROC_BIND=close, OMP_PLACES=threads} \\
 \textbf{Team Size} & \texttt{Kokkos::AUTO()} \\
 \textbf{Base Block Size} & $256^2$ \\
 \textbf{CFL} & $1.0$ \\
 $\boldsymbol{\beta}$ & $1.0$ \\
 $\boldsymbol{\epsilon}$ & $1 \times 10^{-16}$ \\
 \hline 
\end{tabular}
}
\caption{Architecture and code configuration for the numerical experiments conducted on the Intel 18 cluster at Michigan State University's Institute for Cyber-Enabled Research. As with the loop experiments in \cref{subsec: Loop Optimizations}, we encourage the compiler to use AVX-512 instructions and avoid the use of prefetching. All available threads within the node (40 threads/node) were used in the experiments. Each node consists of two Intel Xeon Gold 6148 CPUs and at least 83 GB of memory. We wish to note that hyperthreading is not supported on this system. As mentioned in \cref{subsec: Loop Optimizations}, when hyperthreading is not enabled, \texttt{Kokkos::AUTO()} defaults to a team size of 1. In cases where the base block size did not divide the problem evenly, this parameter was adjusted to ensure that blocks were nearly identical in size. The parameter $\beta$, which does not depend on $\Delta t$ is used in the definition of $\alpha$. For details on the range of admissible $\beta$ values, we refer the reader to \cite{christlieb2019kernel, christlieb2020_NDAD}, where this parameter was introduced. Lastly, recall that $\epsilon$ is the tolerance used in the NN constraints.}
\label{tab:numerical experiment config table}
\end{table}

\subsection{Description of Test Problems and Convergence Experiments}
\label{subsec:Convergence}

\begin{figure}[t]
    \centering
    \includegraphics[width=0.5\linewidth]{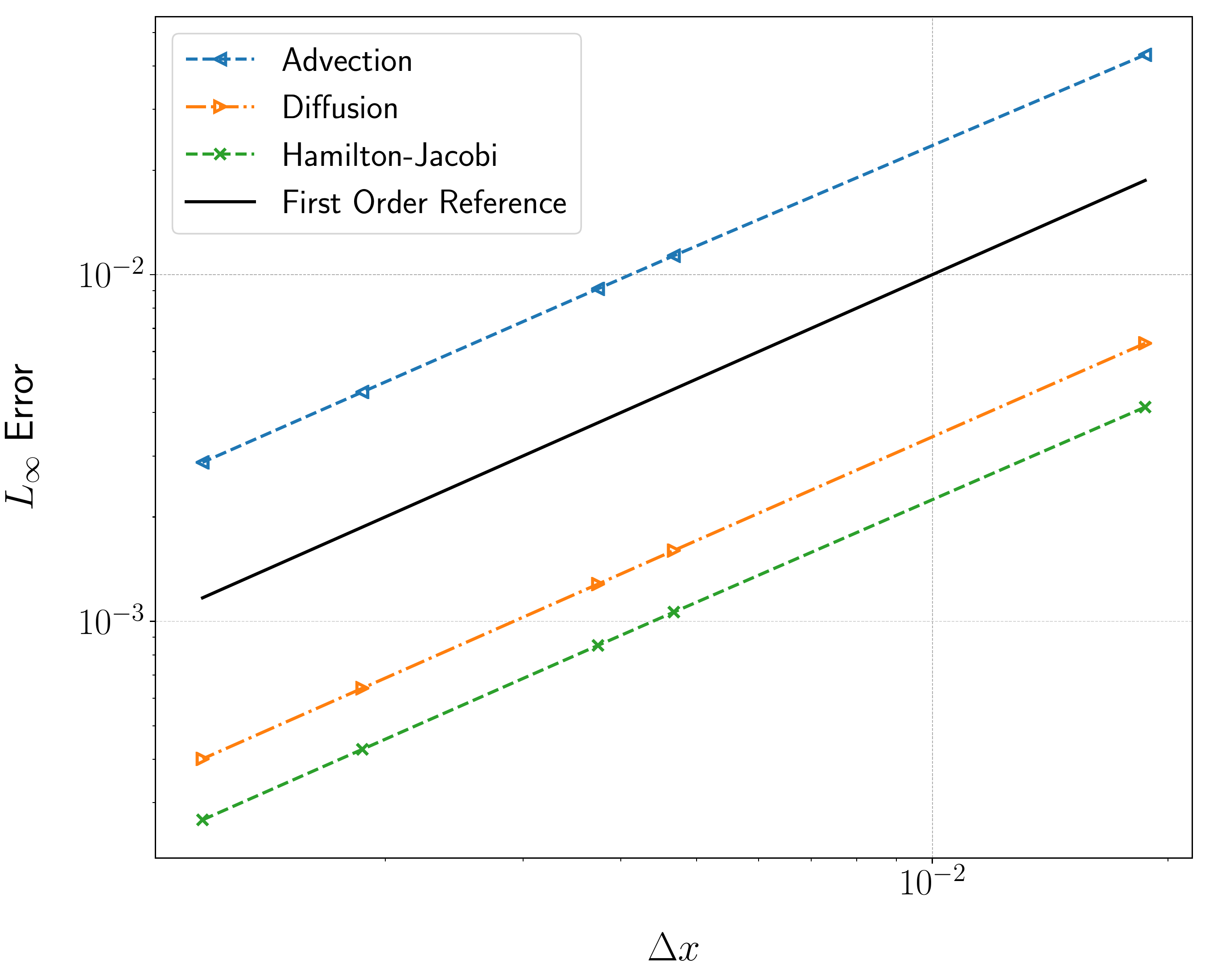}
    \caption{Convergence results for each of the 2-D example problems. Results were obtained using 9 MPI ranks with 40 threads/node. Also included is a first-order reference line (solid black). Our convergence results indicate first-order accuracy resulting from the low-order temporal discretization. The final reported $L_{\infty}$ errors for each of the applications, on a grid containing $5277^2$ total zones, are $2.874 \times 10^{-3}$ (advection), $4.010 \times 10^{-4}$ (diffusion), and $2.674 \times 10^{-4}$ (H-J).} 
    \label{fig:convergence_plot}
\end{figure}

Despite the fact that we are primarily focused on developing codes for high-performance applications, we must also ensure that the parallel algorithms produce reliable answers. Here, we demonstrate convergence of the 2D hybrid parallel algorithms on several test problems, including a nonlinear example that employs the adaptive time stepping rule outlined in \cref{alg:adaptive rule}. The convergence results used 9 nodes, with 40 threads per node, assigning 1 MPI rank to each node, for a total of 360 threads. The quadrature method used to construct the local integrals is the sixth-order WENO-quadrature rule, described in \cref{subsec:Spatial Discretization}, which uses only the linear weights. The numerical solution, in each of the examples, remains smooth over the corresponding time interval of interest. Therefore, it is not necessary to transform the linear quadrature weights to nonlinear ones. According to the analysis of the truncation error presented in \cite{christlieb2019kernel, christlieb2020_NDAD}, retaining a single term in the partial sums for $\mathcal{D}_{*}$ should yield a first-order convergence rate, depending on the choice of $\alpha$. Convergence results for each of the three test problems defined in \cref{ex:Advection example,ex:Diffusion example,ex:HJE example} are provided in \cref{fig:convergence_plot}.

\subsubsection*{Example 1: Linear Advection Equation}
\label{ex:Advection example}

The first test problem considered in this work is the 2D linear advection equation
\begin{align*}
    & \partial_{t} u + \partial_{x} u + \partial_{y} u = 0, \quad (x,y) \in [0,2\pi]^{2}, \\
    &u_{0}(x,y) = \frac{1}{4}(1 - \cos(x))(1 - \cos(y)), \\
    & \text{s.t. two-way periodic BCs}.
\end{align*}
We evolve the numerical solution to the final time $T = 2\pi$. In the experiments, we used the same number of mesh points in both directions, with $\alpha_{x} = \alpha_{y}  = \beta/\Delta t$, with $\beta = 1$. 
% The time step for the simulation was selected as $\Delta t = \Delta x$ (CFL = 1) and was held fixed for the duration of the simulation. The nearest neighbor condition given by \cref{eq:NN-DD constraint 1}, with $\epsilon = 1 \times 10^{-16}$, was used to adjust the time step, if necessary. 
While this problem is rather simple and does not highlight many of the important features of our algorithm, it is nearly identical to the code for a nonlinear example. For initial experiments, a simple test problem is preferable because it gives more control over quantities which are typically dynamic, such as wave speeds. Moreover, the error can be easily computed from the exact solution $$u(x,y,t) = u_0(x - t, y - t). $$ 

% In \cref{subsec:CFL Studies}, we use this feature to study the behavior of the nearest neighbor restriction \eqref{eq:NN-DD constraint 1} by examining the effect of the CFL on the time-to-solution. 

\subsubsection*{Example 2: Linear Diffusion Equation}
\label{ex:Diffusion example}

The next test problem that we consider is the linear diffusion equation
\begin{align*}
    & \partial_{t} u = \partial_{xx} u + \partial_{yy} u, \quad (x,y) \in [0,2\pi]^{2}, \\
    &u_{0}(x,y) = \sin(x) + \sin(y), \\
    & \text{s.t. two-way periodic BCs}.
\end{align*}
The numerical solution is evolved from $(0,T]$, with $T = 1$, in order to prevent substantial decay. As with the previous example, we use an equal number of mesh points in both directions, so that $\Delta x = \Delta y$. The fixed time stepping rule was used, with $\Delta t = \Delta x$. Compared with the previous example, we used the parameter definitions $\alpha_x = \alpha_y = 1/\sqrt{\Delta t}$, which corresponds to $\beta = 1$ in the definition for second derivative operators. The exact solution for this problem is given by
\begin{equation*}
    u(x,y,t) = e^{-t} \Big( \sin(x) + \sin(y) \Big).
\end{equation*}

Characteristically, this example is different from the advection equation in the previous example, which allows us to illustrate some key features of the method. Firstly, code developed for advection operators can be reused to build diffusion operators, an observation made in \cref{subsec:Connections}. More specifically, to construct the left and right-moving local integrals, we used the same linear WENO quadrature as with the advection equation in Example 1. However, we note that this particular example could, instead, use a more compact quadrature to eliminate the halo communication, which would remove a potential synchronization point. The second feature concerns the time-to-solution and is related to the unconditional stability of the method. Linear diffusion equations, when solved by an explicit method, are known to incur a harsh stability restriction on the time step, namely, $\Delta t \sim \Delta x^2$, making long-time simulations prohibitively expensive. The implicit aspect of this method drastically reduces the time-to-solution, as one can now select time steps which are, for example, proportional to the mesh spacing. This benefit is further emphasized by the overall speed of the method, which can be observed in \cref{subsec:Weak Scaling,subsec:Strong Scaling}. 

\subsubsection*{Example 3: Nonlinear Hamilton-Jacobi Equation}
\label{ex:HJE example}

The last test problem we consider is the nonlinear H-J equation
\begin{align*}
    & \partial_{t} u + \frac{1}{2} \left( 1 + \partial_{x} u + \partial_{y} u \right)^{2} = 0, \quad (x,y) \in [0,2\pi]^{2}, \\
    &u_{0}(x,y) = 0, \\
    & \text{s.t. two-way periodic BCs}.
\end{align*}
To prevent the characteristic curves from crossing, which would lead to jumps in the derivatives of the function $u$, the numerical solution is tracked over a short time, i.e., $T = 0.5$. We applied a high-order linear WENO quadrature rule to approximate the left and right-moving local integrals and used the same parameter choices for $\alpha_x$, $\alpha_y$, and $\beta$, as with the advection equation in Example 1. However, since the wave speeds fluctuate based on the behavior of the solution $u$, we allow the time step to vary according to \eqref{eq:Adaptive time rule}, which requires the use of the distributed adaptive time stepping rule outlined in \cref{alg:adaptive rule}.

Typically, an exact solution is not available for such problems. Therefore, to test the converge of the method, we use a manufactured solution given by
\begin{equation*}
    u(x,y,t) = t \Big( \sin(x) + \sin(y) \Big),
\end{equation*}
with a corresponding source term included on the right-hand side of the equation. Methods employed to solve this class of problems are typically explicit, with a shock-capturing method being used to handle the appearance of ``cusps" that would otherwise lead to jumps in the derivative of the solution. A brief summary of such methods is provided in our recent paper \cite{christlieb2020nonuniformHJE}, where extensions of successive convolution were developed for curvilinear and non-uniform grids. The method follows the same structural format as an explicit method with the ability to take larger time steps as in an implicit method. However, the explicit-like structure of this method does not require iteration for nonlinear terms and allows for a more straightforward coupling with high-order shock-capturing methods. We wish to emphasize that despite the fact that this example is nonlinear, the only major mathematical difference with Example 1 is the evaluation of a different Hamiltonian function.

% Change figsize to 0.65 for SIAM...

\subsection{Weak Scaling Experiments}
\label{subsec:Weak Scaling}

% Change figsize to 0.9 for SIAM...
\begin{figure}[!ht]
    \centering
    % Trim left bottom right top
    \subfloat{\label{fig:weak_scaling_49_nodes_fastest}}{\includegraphics[trim=0 95 0 0,clip, width=0.85\linewidth]{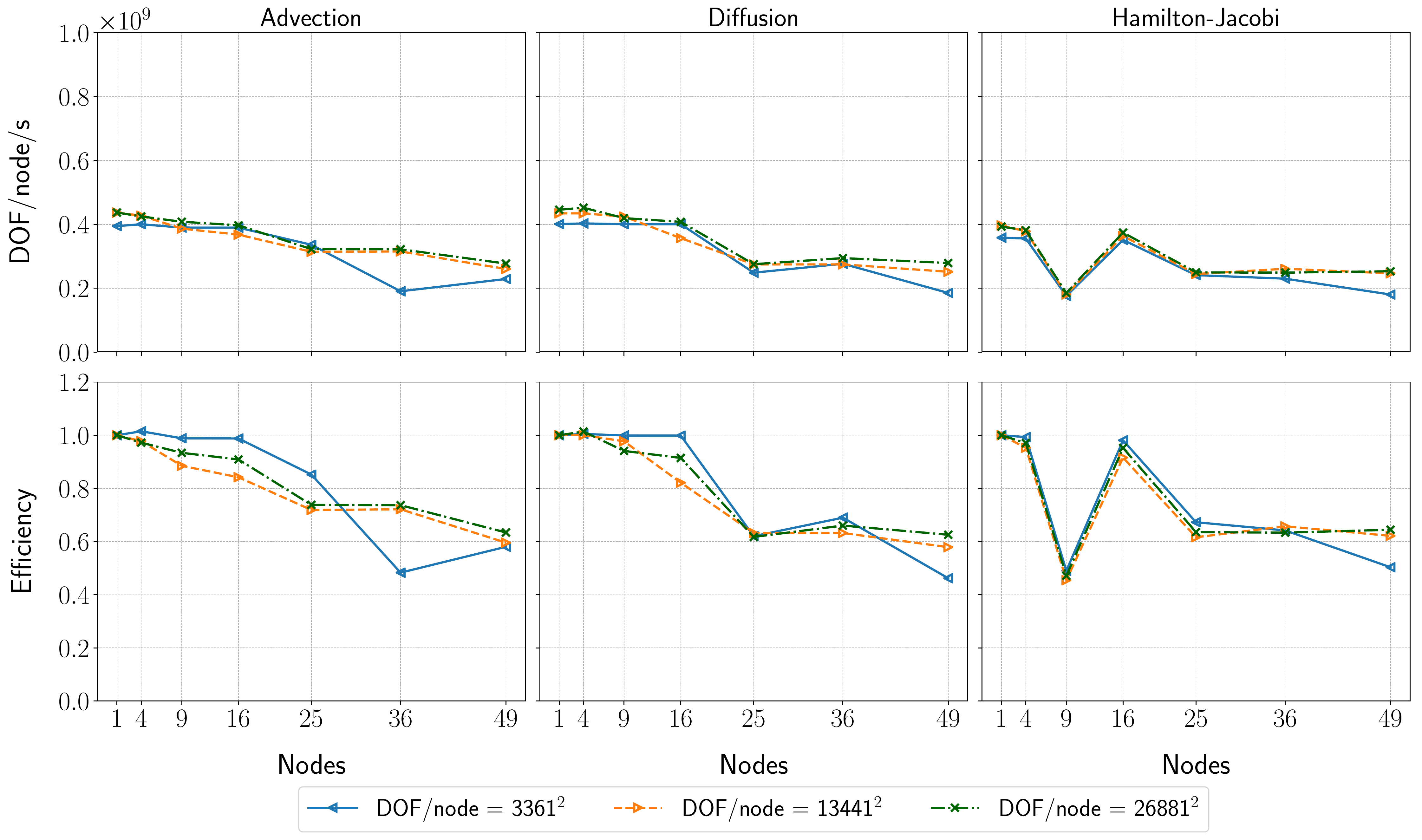}}
    % Trim the legend off of the bottom plot because it is the same as the one above it
    \subfloat{\label{fig:weak_scaling_49_nodes_average}}{\includegraphics[trim=0 0 0 0,clip, width=0.85\linewidth]{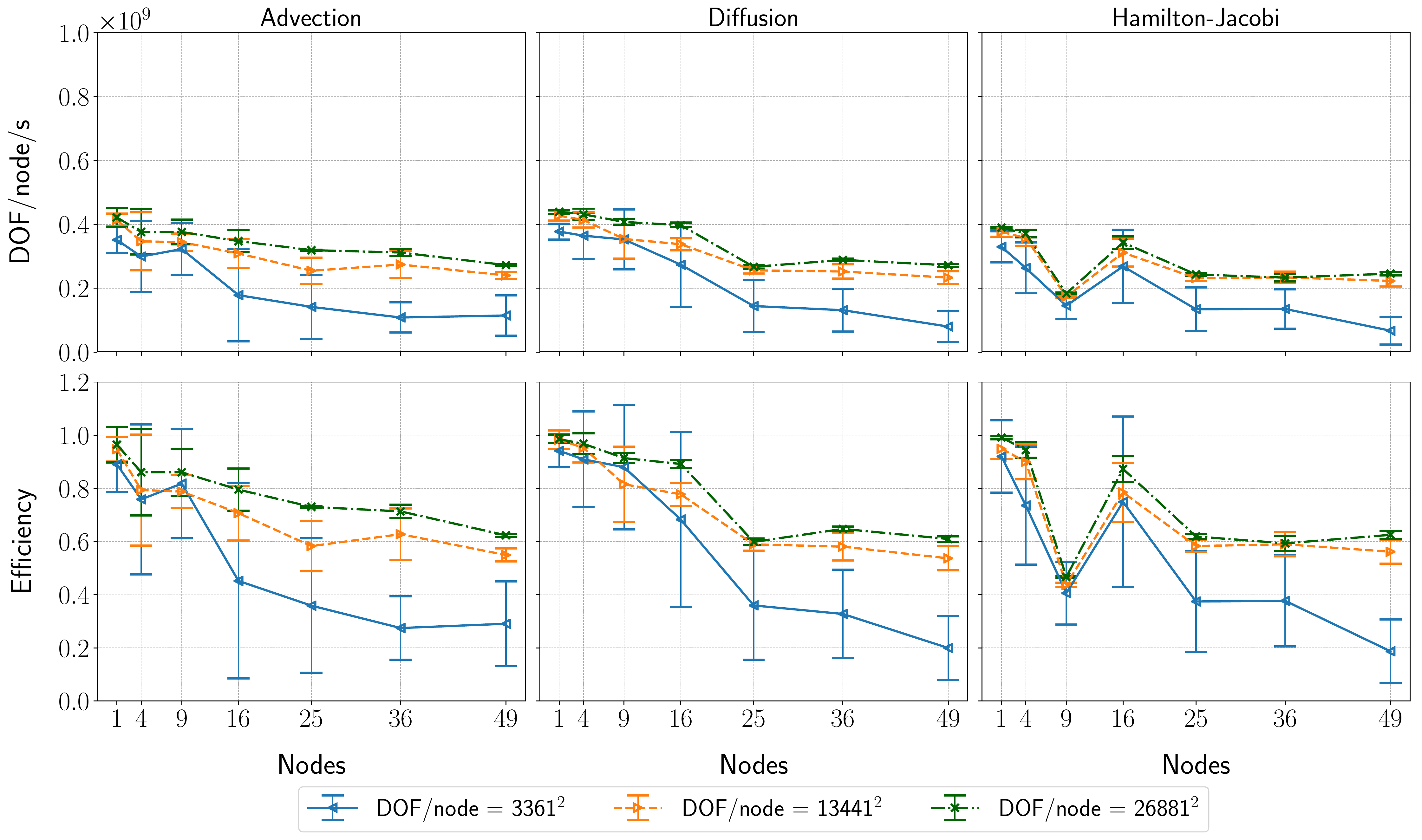}}
    \caption{Weak scaling results, for each of the applications, using up to 49 nodes (1960 cores). For each of the applications, we have provided the update rate and weak scaling efficiency computed via the fastest time/step (top) and average time/step (bottom). Results for advection and diffusion applications is quite similar, despite the use of different operators. The results for the H-J application seem to indicate that no major performance penalties are incurred by use of the adaptive time stepping method. Scalability appears to be excellent, up to 16 nodes (640 cores), then begins to decline. While some loss in performance, due to network effects, is to be expected, this loss appears to be larger than was previously observed. The nodes used in the runs were not contiguous, which hints at a possible sensitivity to data locality.} 
    \label{fig:large_weak_scaling}
\end{figure}

% Change figsize to 0.9 for SIAM...
\begin{figure}[!ht]
    \centering
    \subfloat{\label{fig:weak_scaling_9_nodes_fastest}}{\includegraphics[trim=0 170 0 0,clip, width=0.85\linewidth]{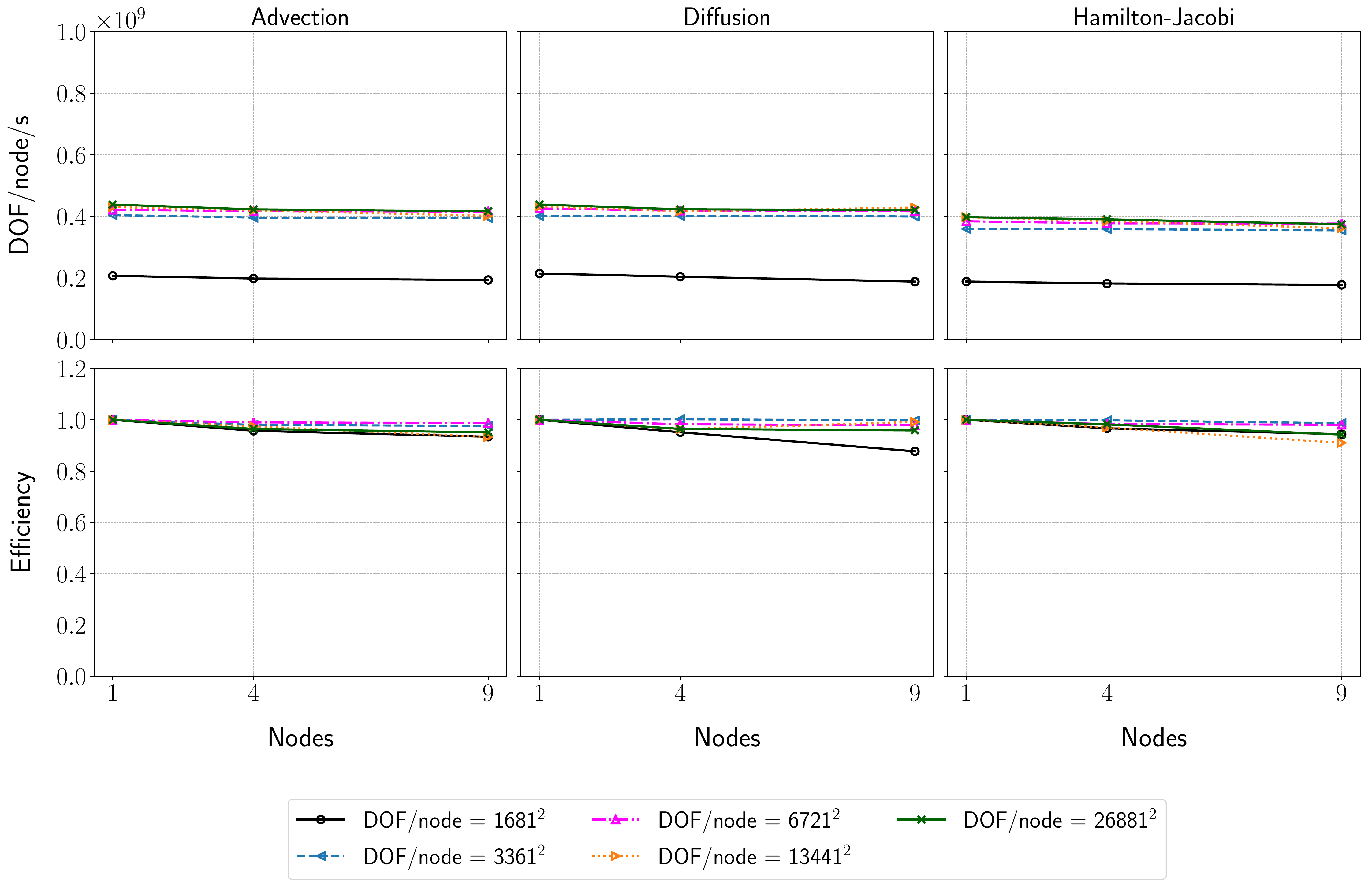}}
    \subfloat{\label{fig:weak_scaling_9_nodes_average}}{\includegraphics[trim=0 0 0 0 0,clip, width=0.85\linewidth]{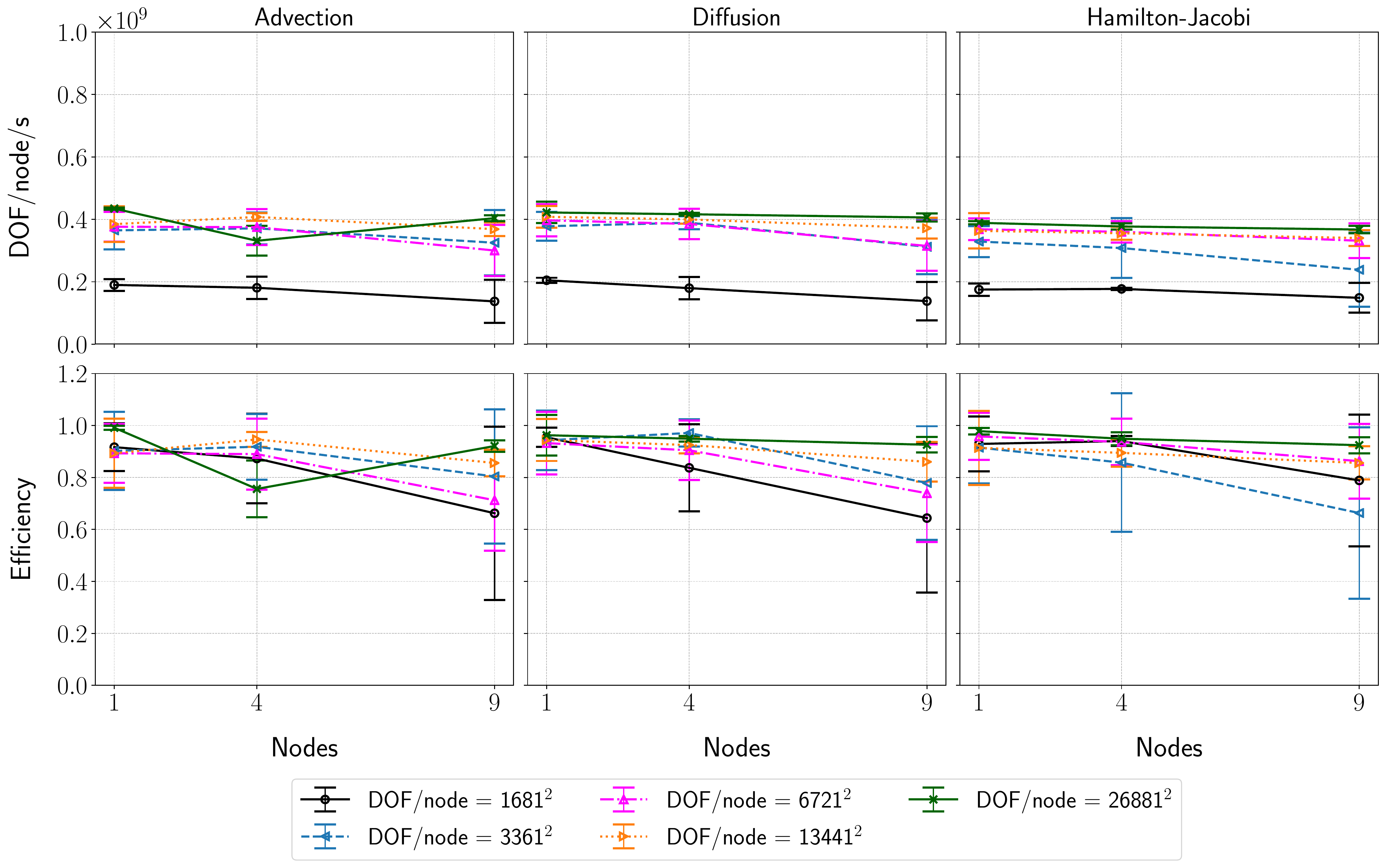}}
    \caption{Weak scaling results obtained with contiguous allocations of up to 9 nodes (360 cores) for each of the applications. For comparison, the same information is displayed as in \cref{fig:large_weak_scaling}. Data from the fastest trials indicates nearly perfect weak scaling, across all applications, up to 9 nodes, with a consistent update rate between $2 - 4 \times 10^8$ DOF/node/s. A comparison of the fastest timings between the large and small runs supports our claim that data proximity is crucial to achieving the peak performance of the code. Note that size the error bars are generally smaller than those in \cref{fig:large_weak_scaling}. This indicates that the timing data collected from individual trials exhibits less overall variation.}
    \label{fig:small_weak_scaling}
\end{figure}

A useful performance property for examining the scalability of parallel algorithms describes how they behave when the compute resources are chosen proportionally to the size of the problem. Here, the amount of work per compute unit remains fixed, and the compute units are allowed to fluctuate. Weak scaling assumes ideal or best-case performance for the parallel components of algorithms and ignores the influence of bottlenecks imposed by the sequential components of a code. Therefore, for $N$ compute units, we shall expect a speedup of $N$. This motivates the following definitions for speedup and efficiency in the context of weak scaling:
\begin{equation*}
    S_N = \frac{N T_1}{T_N}, \quad E_N = \frac{S_N}{N} \equiv \frac{T_1}{T_N}.
\end{equation*}
Therefore, with weak scaling, ideal performance is achieved when the run times for a fixed work size (or, equivalently, the DOF/node/s) remain constant, as we vary the compute units. To scale the problem size, we take advantage of the periodicity for the test problems. The base problem on $[0,2\pi] \times [0,2\pi]$ can be replicated across nodes so that the total work per node remains constant. Provided in \cref{fig:large_weak_scaling} are plots of the weak scaling data --- specifically the update metric \cref{eq:performance metric def} and the corresponding efficiency --- obtained from the fastest of 10 trials of each configuration, using up to 49 nodes (1,960 cores). 

These results generally indicate good performance, both in terms of the update frequency and efficiency, for a variety of problem sizes. Weak scalability appears to be excellent up to 16 nodes (640 cores), then begins to decline, most likely due to network effects. The performance behavior for advection and diffusion applications is quite similar, which is to be expected, since the parallel algorithms used to construct the base operators are nearly identical. With regard to the Hamilton-Jacobi application, we see that the performance is similar to the other applications at larger node counts. This seems to indicate that no major communication penalties are incurred by use of the adaptive time stepping method shown in \cref{alg:adaptive rule}, compared to fixed time stepping. Additionally, in the Hamilton-Jacobi application, we observe a sharp decline in the performance at 9 nodes in \cref{fig:large_weak_scaling}. A closer investigation reveals that this is likely an artifact of the job scheduler for the system on which the experiments were conducted, as we were unable to secure a ``contiguous" allocation of nodes. This has the unfortunate consequence of not being able to guarantee that data for a particular trial remain in close \textit{physical} proximity. This could result in issues such as network contention and delays that exacerbate the cost of communication relative to the computation as discussed in \cref{sec:Implementation}. An non-contiguous placement of data is problematic for codes with inexpensive operations, such as the methods shown here, because the work may be insufficient to hide this increased cost of communication. For this reason, we chose to include plots containing the averaged weak scaling data in \cref{fig:large_weak_scaling}, which contains error bars calculated from the sample standard deviation. The noticeable size of the error bars in these plots generally indicates a large degree of variation in the timings collected from trials.

To more closely examine the importance of data proximity on the nodes, we repeated the weak scaling study, but with node counts for which a contiguous allocation count could be guaranteed. We have provided results for the fastest and averaged data in \cref{fig:small_weak_scaling}. Data collected from the fastest trials indicates nearly perfect weak scaling, across all applications, up to 9 nodes, with a consistent update rate between $2 - 4 \times 10^8$ DOF/node/s. For convenience, these results were plotted with the same markers and formats so that results from the larger experiments in \cref{fig:large_weak_scaling} could be compared directly. A comparison of the fastest timings between the large and small runs supports our claim that data proximity is crucial to achieving the peak performance of the code. Furthermore, the error bars for the contiguous experiments displayed in \cref{fig:small_weak_scaling} show that the individual trials exhibit less overall variation in timings. 

\subsection{Strong Scaling Experiments}
\label{subsec:Strong Scaling}

% Change figsize to 0.9 for SIAM...
\begin{figure}[!ht]
    \centering
    \subfloat{\label{fig:strong_scaling_9_nodes_fastest}}{\includegraphics[trim=0 130 0 0,clip, width=0.85\linewidth]{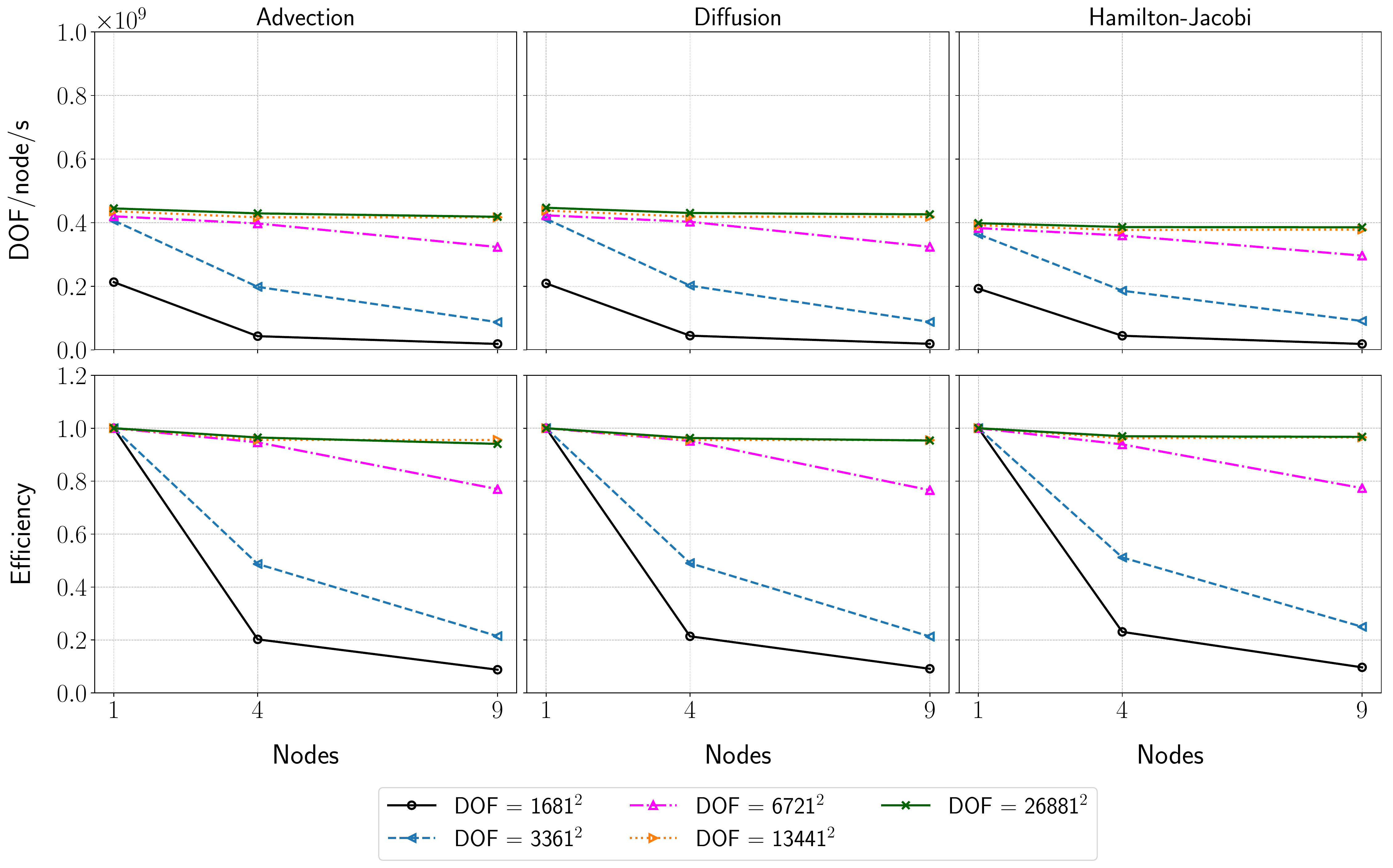}}
    \subfloat{\label{fig:strong_scaling_9_nodes_average}}{\includegraphics[trim=0 0 0 0,clip, width=0.85\linewidth]{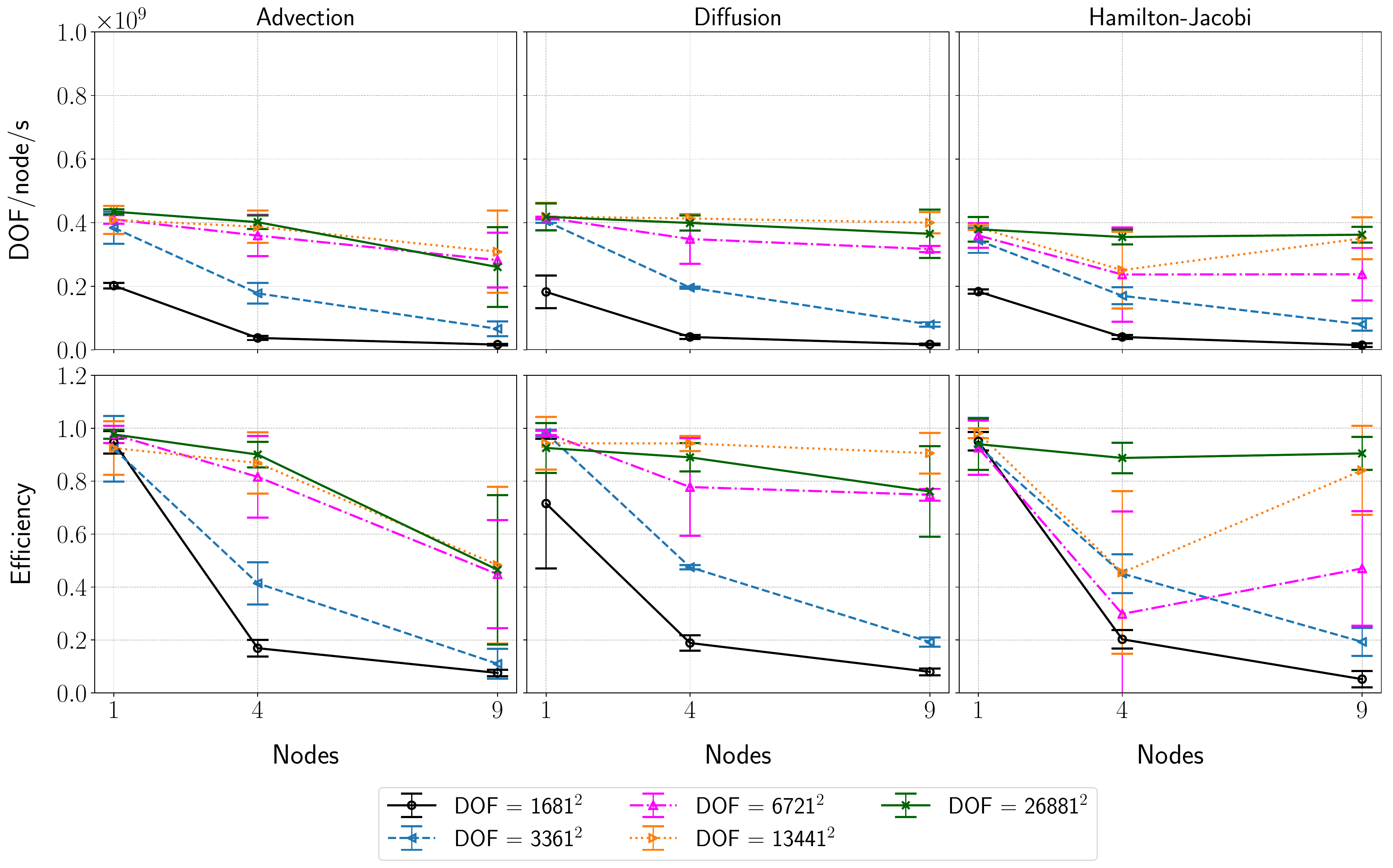}}
    \caption{Strong scaling results for each of the applications obtained on contiguous allocations of up to 9 nodes (360 cores). Displayed among each of the applications are the update rate and strong scaling efficiency computed from the fastest time/step (top) and average time/step (bottom). This method does not contain a substantial amount of work, so we do not expect good performance for smaller base problem sizes, as the work per node becomes insufficient to hide the cost of communication. Larger base problem sizes, which introduce more work, are capable of saturating the resources, but will at some point become insufficient. Moreover, threads become idle when the work per node fails to introduce enough blocks.}
    \label{fig:small_strong_scaling}
\end{figure}

Another form of scalability considers a fixed problem size and examines the effect of varying the number of work units used to find the solution. In these experiments, we allow the work per compute unit to decrease, which helps identify regimes where sequential bottlenecks in algorithms become problematic, provided we are granted enough resources. Applications which are said to strong scale exhibit run times which are inversely proportional to the number of resources used. For example, when $N$ compute units are applied to a problem, one expects the run to be $N$ times faster than with a single compute unit. Additionally, if an algorithm's performance is memory bound, rather than compute bound, this will, at some point, become apparent in these experiments. Supplying additional compute units should not improve performance, if more time is spent fetching data, rather than performing useful computations. This motivates the following definitions for speedup and efficiency in the context of strong scaling:
\begin{equation*}
    S_N = \frac{T_1}{T_N}, \quad E_N = \frac{S_N}{N}. 
\end{equation*}
Here, $N$ is the number of nodes used, so that $T_{1}$ and $T_{N}$ correspond to the time measured using a single node and N nodes, respectively. Results of our strong scaling experiments are provided in \cref{fig:small_strong_scaling}. As with the weak scaling experiments, we have plotted the update metric \cref{eq:performance metric def} along with the strong scaling efficiency using both the fastest and averaged configuration data from a set of 10 trials. In contrast to weak scaling, strong scaling does not assume ideal speedup, so one could plot this information; however, the information can be ascertained from the efficiency data, so we refrain from plotting this data. 

Results from these experiments show decent strong scalability for the N-N method. This method does not contain a substantial amount of work, so we do not expect good performance for smaller base problem sizes, as the work per node becomes insufficient to hide the cost of communication. On the other hand, larger base problem sizes, which introduce more work, are capable of saturating the resources, but will at some point become insufficient. This behavior is apparent in our efficiency plots. Increasing the problem size generally results in an improvement of the efficiency and speedup for the method. Part of these problems can be attributed to the use of a blocking pattern for loops structures discussed in \cref{subsec: Shared Memory Algorithms}. Depending on the size of the mesh, it may be the case that the block size and the team size set by the user result in idle threads. One possible improvement is to simply increase the team size so that there are fewer idle threads within an MPI task. Alternatively, one can adjust the number of threads per task, so that each task is responsible for fewer threads. While these approaches can be implemented with no changes to the code, they will likely not resolve this issue. Profiling seems to indicate that the source of the problem is the low arithmetic intensity of the reconstruction algorithms. In other words, the method is memory bound because the calculations required in the reconstructions are inexpensive relative to the cost of retrieving data from memory. As part of our future work, we plan to investigate such limitations through the use of detailed roofline models.  

\subsection{Effect of CFL}
\label{subsec:CFL Studies}

In order to enforce a N-N dependency for our domain decomposition algorithm, we obtained several possible restrictions on $\Delta t$, depending on the problem and the choice of $\alpha$. In the case of linear advection, we would, for example, require that
\begin{equation*}
    \Delta t \leq -\frac{\beta L_{m}}{ c_{\text{max}} \log(\epsilon)}.
\end{equation*}
with the largest possible time step permitting N-N dependencies being set by the equality. Admittedly, such a restriction is undesirable. As mentioned in \cref{subsec: NN criterion}, this assumption can be problematic if the problem admits fast waves ($c_{\text{max}}$ is large) and/or if the block sizes are particularly small ($L_m$ is small). In many applications, the former circumstance is quite common. However, our test problem contains fixed wave speeds so this is less of an issue. The latter condition is a concern for configurations which use many blocks, such as a large simulation on many nodes of a cluster. Another potential circumstance is related to the granularity of the blocks. For example, in these experiments, we use 1 MPI rank per compute node. However, it may be advantageous to consider different task configurations, e.g., using 1 (or more) rank(s) per NUMA region of a compute node.  

% Change figsize to 0.85 for SIAM...
\begin{figure}[t]
    \centering
    \includegraphics[width=0.75\linewidth]{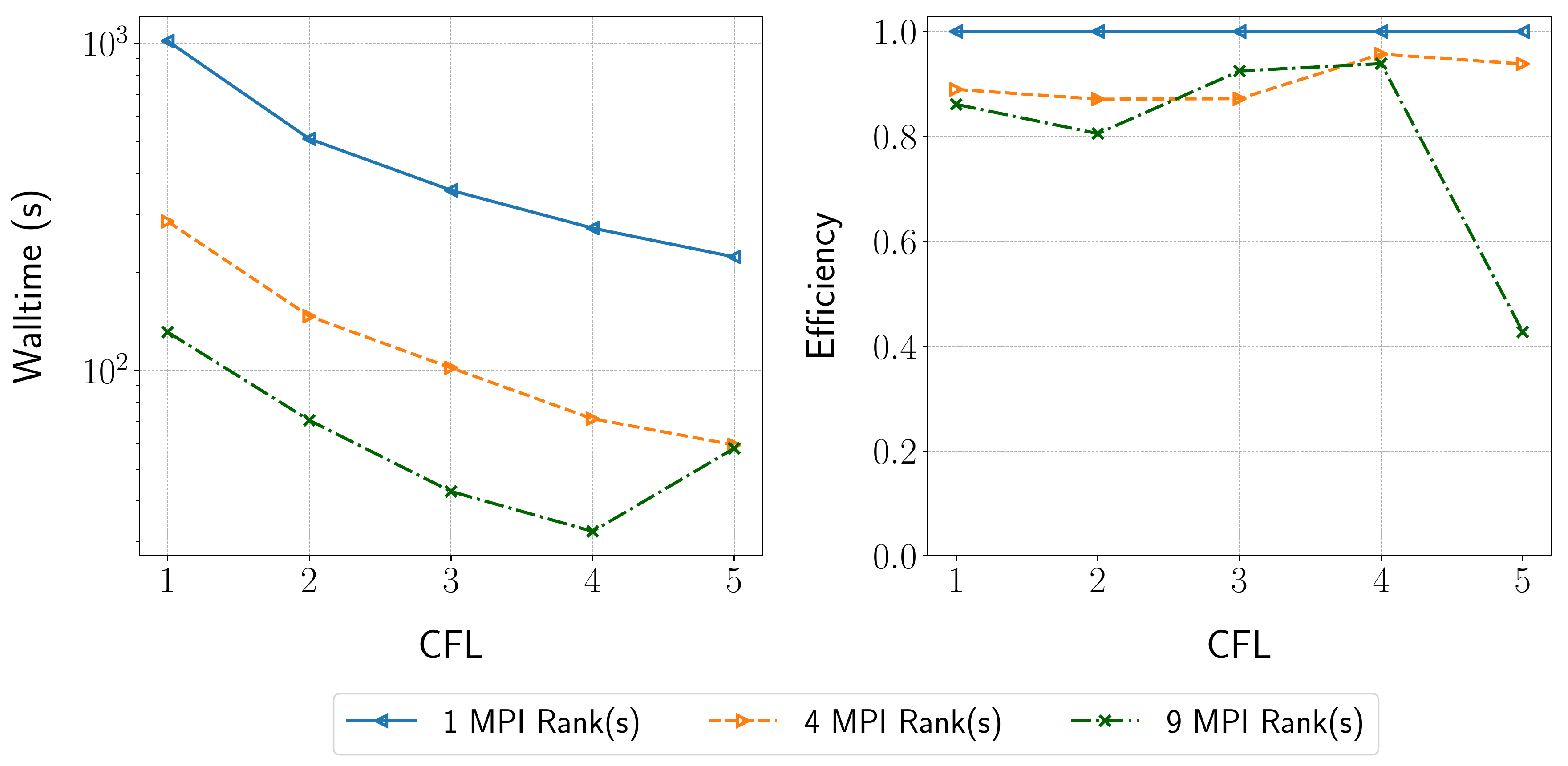}
    \caption{Results on the N-N method for the linear advection equation using a fixed mesh with $5377^{2}$ total DOF and a variable CFL number. In each case, we used the fastest time-to-solution collected from repeating each configuration a total of 20 times. This particular data was collected using an older version of the code, compiled with GCC, which did not use the blocking approach. For larger block sizes, increasing the CFL has a noticeable improvement on the run time, but as the block sizes become smaller, the gains diminish. For example, if 9 MPI ranks are used, improvements are observed as long as $\text{CFL} \leq 4$. However, when $\text{CFL} = 5$, the run times begin to increase, with a significant decrease in efficiency. As the blocks become smaller, $\Delta t$ needs to be adjusted (decreased) so that the support of the non-local convolution data not extend beyond N-Ns.} 
    \label{fig:CFL_study}
\end{figure}

A larger CFL parameter is generally preferable because it reduces the overall time-to-solution. Eventually, however, for a given CFL, there will be a crossover point, where the time step restriction causes the performance to drop due to the increasing number of \textit{sequential} time steps. This experiment used a highly refined grid and varied the CFL number, using up to 9 nodes. Results from the CFL experiments are provided in \cref{fig:CFL_study}. The data was obtained using an older version of our parallel algorithms, compiled with GCC 8.2.0-2.31.1, which does not use blocking. By plotting the behavior according to the number of nodes (ranks) used, we can fix the lengths of the blocks, hence $L_m$, and change the time step to identify the breakdown region. We observe a substantial decrease in performance for the 9 node configuration, specifically when the CFL number increases from 4 to 5. For more complex problems with dynamic wave behavior, this breakdown may be observed earlier. In response to this behavior, a user could simply increase or relax the tolerance, but the logarithm tends to suppress the impact of large relaxations. Another option, which shall be considered in future work is to include more information from neighboring ranks by either eliminating this condition or, at the least, communicating enough information to achieve a prescribed tolerance. 

\section{Conclusion}
\label{sec:conclusion}

% Reiterate the essential pieces of the paper (What did I talk about?)
In this paper, we presented hybrid parallel algorithms capable of addressing a wide class of both linear and nonlinear PDEs. To enable parallel simulations on distributed systems, we derived a set of conditions that use available wave speed (and/or diffusivity) information, along with the size of the sub-domains, to limit the communication through an adjustment of the time step. Although not considered here, these conditions, which are needed to ensure accuracy, rather than the stability of the method, can be removed at the cost of additional communication. Using these restrictions, boundary conditions could be enforced across sub-domains in the decomposition. Results were obtained for 2-D examples consisting of linear advection, linear diffusion, and a nonlinear H-J equation to highlight the versatility of the methods in addressing characteristically different PDEs. 

As part of the implementation, we used constructs from the Kokkos performance portability library to parallelize the shared memory components of the successive convolution algorithms. We extracted essential loop structures from the algorithms and analyzed a variety of parallel execution policies in an effort to develop an efficient application. These experiments considered several common optimization techniques, such as vectorization, cache-blocking, and placement of threads. From these experiments, we chose to use a blocked iteration pattern in which threads (or teams thereof) are mapped to blocks of an array with vector instructions being applied to 1-D line segments. These design choices offer a large degree of flexibility, which is an important consideration as we proceed with experimentation on other architectures, including GPUs, to leverage the full capabilities provided by Kokkos. By exploiting the stability properties of the representations, we also developed an adaptive time stepping method for distributed systems that uses ``lagged" wave speed information to calculate the time step. While the methods presented here do not require adaptive time stepping for stability, it was included as an option because of its ability to prevent excessive numerical diffusion, as observed in previous work.

Convergence and scaling properties for the hybrid algorithms were established using at most 49 nodes (1960 cores), with a peak performance $ > 10^8$ DOF/node/s. Larger weak scaling experiments, which used up to 49 nodes (1960 cores), initially performed reasonably well, with all applications later tending to 60\% efficiency corresponding (roughly) to $2 \times 10^8 $ DOF/node/s. While some performance loss is to be expected from network related complications, we found this to be much larger than what was observed in prior experiments. Later, it was discovered that the request for a contiguous allocation could not be accommodated so data locality in the experiments was compromised. By repeating the experiments on a smaller collection of nodes, which granted this request, we discovered that data locality plays a pivotal role in the overall performance of the method. We observe that a large base problem size is required to achieve good strong scaling. Furthermore, when threads are prescribed work at a coarse granularity (i.e., across blocks, rather than entries within the blocks), one must ensure that the problem size is capable of saturating the resources to avoid idle threads. This approach introduces further complications for strong scaling, as the workload per node drops substantially while the block size remains fixed. Finite difference methods which, generally, do not generate a substantial amount of work, are quite similar to successive convolution.  Therefore, we do not expect excellent strong scalability. Certain aspects of the algorithms can be tuned to improve the arithmetic intensity, which will improve the strong scaling behavior. At some point, however, the algorithms will be limited by the speed of memory transfers rather than computation. We also provided experimental results that demonstrate the limitations of the N-N condition in the context of strong scaling.  

While we have presented several new ideas with this work, there is still much untapped potential with successive convolution methods. Firstly, optimizations on GPU architectures, which shall play an integral role in the upcoming exascale era, need to be explored and compared with CPUs. A roofline model should be developed for these algorithms to help identify key limitations and bottlenecks and formulate possible solutions. Although this work considered only first-order time discretizations, our future developments shall be concerned with evaluating a variety of high-order time discretization techniques in an effort to increase the efficiency of the method. Lastly, the parallel algorithms should be modified to enable the possibility of mesh adaptivity, which is a common feature offered by many state-of-the-art computing libraries.

\appendix

\section{ Example for Linear Advection}
\label{subsec:MOLT for Advection}

Suppose we wish to solve the 1D linear advection equation:
\begin{equation}
    \label{eq:1D advection}
    \partial_{t} u + c \partial_{x} u = 0, \quad (x,t) \in (a,b) \times \mathbb{R}^{+},
\end{equation}
where $c > 0$ is the wave speed and leave the boundary conditions unspecified. The procedure for $c < 0$ is analogous. Discretizing \eqref{eq:1D advection} in time with backwards Euler yields a semi-discrete equation of the form
\begin{equation*}
    \frac{u^{n+1}(x) - u^{n}(x)}{\Delta t} + c \partial_{x} u^{n+1}(x) = 0.
\end{equation*}
If we rearrange this, we obtain a linear equation of the form
\begin{equation}
    \label{eq:1D semidiscrete advection}
    \mathcal{L}[u^{n+1}; \alpha](x) = u^{n}(x),
\end{equation}
where we have used
\begin{equation*}
    \alpha := \frac{1}{c\Delta t},\quad \mathcal{L}:= \mathcal{I} + \frac{1}{\alpha} \partial_{x}.    
\end{equation*}
By reversing the order in which the discretization is performed, we have created a sequence of BVPs at discrete time levels. If we had discretized equation \eqref{eq:1D advection} using the MOL formalism, then $\mathcal{L}$ would be an algebraic operator. To solve equation \eqref{eq:1D semidiscrete advection} for $u^{n+1}$, we analytically invert the operator $\mathcal{L}$. Notice that this equation is actually an ODE, which is linear, so the problem can be solved using methods developed for ODEs. If we apply the integrating factor method to the problem, we obtain
\begin{equation*}
    \partial_{x} \left[ e^{\alpha x} u^{n+1}(x) \right] = \alpha e^{\alpha x} u^{n}(x).
\end{equation*}
To integrate this equation, we use the fact that characteristics move to the right, so integration is performed from $a$ to $x$. After rearranging the result, we arrive at the update equation
\begin{align*}
    u^{n+1}(x) &= e^{-\alpha(x-a)}  u^{n+1}(a) + \alpha \int_{a}^{x} e^{-\alpha(x - s)} u^{n}(s) \, ds, \\ 
    &\equiv e^{-\alpha(x-a)} A^{n+1} + \alpha \int_{a}^{x} e^{-\alpha(x - s)} u^{n}(s) \, ds, \\
    &\equiv  \mathcal{L}^{-1}[u^{n}; \alpha](x).
\end{align*}
This update displays the origins of the implicit behavior of the method. While convolutions are performed on data from the previous time step, the boundary terms are taken at time level $n+1$.

Now that we have obtained the update equation, we need to apply the boundary conditions. Clearly, if the problem specifies a Dirchlet boundary condition at $x = a$, then $A^{n+1} = u^{n+1}(a)$. We can compute a variety of boundary conditions using the update equation
\begin{equation*}
    u^{n+1}(x) = e^{-\alpha(x-a)}  A^{n+1} + \alpha \int_{a}^{x} e^{-\alpha(x - s)} u^{n}(s) \, ds,
\end{equation*}
where
\begin{equation*}
    I[u^{n};\alpha](x) = \alpha \int_{a}^{x} e^{-\alpha(x - s)} u^{n}(s) \, ds.
\end{equation*}
For example, with periodic boundary conditions, we would need to satisfy
\begin{align}
    \label{eq:1D advection periodic value}
    u^{n+1}(a) &= u^{n+1}(b), \\
    \label{eq:1D advection periodic derivative}
    \partial_{x}u^{n+1}(a) &= \partial_{x} u^{n+1}(b).
\end{align}
Applying condition \eqref{eq:1D advection periodic value}, we find that
\begin{equation*}
    A^{n+1} = e^{-\alpha(b-a)} A^{n+1} + \alpha \int_{a}^{b} e^{-\alpha(b - s)} u^{n}(s) \, ds.
\end{equation*}
Solving this equation for $A^{n+1}$ shows that
\begin{equation*}
    A^{n+1} = \frac{I[u^{n};\alpha](b)}{1 - \mu},
\end{equation*}
with $\mu = e^{-\alpha(b - a)}$. Alternatively, we could have started with \eqref{eq:1D advection periodic derivative}, which would give an identical solution. 
% Another boundary condition we might encounter involves information of the derivative at the boundary, i.e., the Neumann condition:
% \begin{equation}
%     \label{eq:1D advection Neumann BC}
%     \partial_{x} u(a) = w(t).
% \end{equation}
% Here, $w(t)$ is some known function supplied by the problem. To apply this condition, we first compute $\partial_{x} \mathcal{L}^{-1}[u;\alpha](x)$:
% \begin{equation*}
%     \partial_{x} \mathcal{L}^{-1}[u;\alpha](x) = -\alpha A^{n+1}e^{-\alpha(x - a)} - \alpha I[u^{n};\alpha](x).
% \end{equation*}
% Evaluating this expression at $x = a$ and using the data in \eqref{eq:1D advection Neumann BC}, we see that
% \begin{equation*}
%     A^{n+1} = -\frac{1}{\alpha} w(t_{n+1}).
% \end{equation*}
While this particular procedure is only applicable to linear problems, this exercise motivates some of the choices made to define operators in the method.

\section{Kokkos Kernels}
\label{sec:kokkos kernels}

This section provides listings, which outline the general format of the Kokkos kernels used in this work. Specifically, we provide structures for the tiled/blocked algorithms (\hyperref[team_tiling]{Listing~\ref*{team_tiling}}) in addition to the kernel that executes the fast summation method along a line (\hyperref[serial_FC]{Listing~\ref*{serial_FC}}). 

% Code listing for the team based patterns
\begin{minipage}{0.90\linewidth}
\begin{lstlisting}[language = C++, caption = An example of coarse-grained parallel nested loop structure.,label=team_tiling]
// Distribute tiles of the array to teams of threads dynamically
Kokkos::parallel_for("team loop over tiles", team_policy(total_tiles, Kokkos::AUTO()), 
    KOKKOS_LAMBDA(team_type &team_member)
{ 
    // Determine the flattened tile index via the team rank
    // and compute the unflattened indices of the tile T_{i,j}
    const int tile_idx = team_member.league_rank();
    const int tj = tile_idx % num_tiles_x;
    const int ti = tile_idx / num_tiles_x;
    
    // Retrieve tile sizes & offsets and
    // obtain subviews of the relevant grid data on tile T_{i,j}
    // ...
    
    // Use a team's thread range over the lines
    Kokkos::parallel_for(Kokkos::TeamThreadRange<>(team_member, Ny_tile), [&](const int iy)
    {
        // Slice to extract a subview of my line's data and
        // call line methods which use vector loops
        // ...
    }
});
\end{lstlisting}
\end{minipage}

\begin{minipage}{0.9\linewidth}
\begin{lstlisting}[language = C++, caption = Kokkos kernel for the fast-convolution algorithm.,label=serial_FC]
// Distribute the threads to lines
Kokkos::parallel_for("Fast sweeps along x", range_policy(0, Ny),
    KOKKOS_LAMBDA(const int iy)
{
    // Slice to obtain the local integrals to which we apply 
    // the convolution kernel to the entire line 
    // ...
});
\end{lstlisting}
\end{minipage}

\section{Sixth-Order WENO Quadrature}

\label{sec:WENO quad info}

We provide the various expressions for the coefficients and smoothness indicators used in the reconstruction process for $J_{R}^{(r)}$. Defining $\nu \equiv \alpha \Delta x$, the coefficients for the fixed stencils are given in \cite{christlieb2020_NDAD} as follows:
\begin{align*}
    c_{-3}^{(0)} &= \frac{6 - 6\nu + 2\nu^{2} - (6-\nu^{2})e^{-\nu} }{6\nu^{3}}, \\
    c_{-2}^{(0)} &= -\frac{6 - 8\nu + 3\nu^{2} - (6 - 2\nu -2\nu^{2})e^{-\nu} }{2\nu^{3}}, \\
    c_{-1}^{(0)} &= \frac{6 - 10\nu + 6\nu^{2} - (6 - 4\nu -\nu^{2} + 2\nu^{2})e^{-\nu} }{2\nu^{3}}, \\
    c_{0}^{(0)}  &= -\frac{6 - 12\nu + 11\nu^{2} - 6\nu^{3} - (6 - 6\nu + 2\nu^{2})e^{-\nu} }{6\nu^{3}},  \\
    &~\\
    c_{-2}^{(1)} &= \frac{6 - \nu^{2} - (6 + 6\nu + 2\nu^{2})e^{-\nu} }{6\nu^{3}}, \\
    c_{-1}^{(1)} &= -\frac{6 - 2\nu - 2\nu^{2} - (6 + 4\nu - \nu^{2} - 2\nu^{3})e^{-\nu} }{2\nu^{3}}, \\
    c_{0}^{(1)}  &= \frac{6 - 4\nu - \nu^{2} + 2\nu^{3} - (6 + 2\nu - 2\nu^{2} )e^{-\nu} }{2\nu^{3}}, \\
    c_{1}^{(1)}  &= -\frac{6 - 6\nu + 2\nu^{2} - (6-\nu^{2})e^{-\nu} }{6\nu^{3}}, \\
    &~\\
    c_{-1}^{(2)} &= \frac{6 + 6\nu + 2\nu^{2} - (6 + 12\nu + 11\nu^{2} + 6\nu^{3} )e^{-\nu} }{6\nu^{3}}, \\
    c_{0}^{(2)}  &= -\frac{6 + 4\nu - \nu^{2} - 2\nu^{3} - (6 + 10\nu + 6\nu^{2})e^{-\nu} }{2\nu^{3}}, \\
    c_{1}^{(2)}  &= \frac{6 + 2\nu - 2\nu^{2} - (6 + 8\nu + 3\nu^{2})e^{-\nu} }{2\nu^{3}}, \\
    c_{2}^{(2)}  &= -\frac{6 - \nu^{2} - (6 + 6\nu + 2\nu^{2})e^{-\nu} }{6\nu^{3}}. 
\end{align*}
The corresponding linear weights are
\begin{align*}
    d_{0} &= \frac{6 - \nu^{2} - (6 + 6\nu + 2\nu^{2})e^{-\nu} }{3\nu ( 2 - \nu - (2 + \nu)e^{-\nu} ) }, \\
    d_{2} &= \frac{ 60 - 60\nu + 15\nu^{2} + 5\nu^{3} - 3\nu^{4} - (60 - 15\nu^{2} + 2\nu^{4})e^{-\nu} }{ 10\nu^{2} (6 - \nu^{2} - (6 + 6\nu + 2\nu^{2})e^{-\nu} )  }, \\
    d_{1} &= 1 - d_{0} - d_{2}.
\end{align*}
% The expressions for the smoothness indicators are given in \cite{christlieb2019kernel} as
% \begin{align*}
%     \beta_{0} &= \frac{13}{12} \left( -v_{i-3} + 3v_{i-2} - 3v_{i-1} + v_{i} \right)^{2} + \frac{1}{4} \left( v_{i-3} - 5v_{i-2} + 7v_{i-1} - 3v_{i} \right)^{2}, \\
%     \beta_{1} &= \frac{13}{12} \left( -v_{i-2} + 3v_{i-1} - 3v_{i} + v_{i+1} \right)^{2} + \frac{1}{4} \left( v_{i-2} - v_{i-1} - v_{i} + v_{i+1} \right)^{2}, \\
%     \beta_{2} &= \frac{13}{12} \left( -v_{i-1} + 3v_{i} - 3v_{i+1} + v_{i+2} \right)^{2} + \frac{1}{4} \left( -3v_{i-1} + 7v_{i} - 5v_{i+1} + v_{i+2} \right)^{2}. 
% \end{align*}
To obtain the analogous expressions for $J_{L}^{(r)}$, we exploit the ``mirror-symmetry" property of WENO reconstructions. That is, one can keep the left side of each of the expressions, then reverse the order of the expressions on the right. Expressions for calculating one particular smoothness indicator, if interested, can be found in \cite{christlieb2020_NDAD}.

%Information regarding the calculation of one example of a smoothness indicator is provided in \cite{christlieb2020_NDAD}.

\section*{Acknowledgments}

The research of the authors was supported partly through computational resources made available by Michigan State University's Institute for Cyber-Enabled Research. Funding for this work was provided in part by AFOSR grants FA9550-19-1-0281 and FA9550-17-1-0394 and NSF grant DMS 1912183. The authors would like to thank Brian O’Shea and Philipp Grete at Michigan State University for helpful discussions relating to Kokkos and some suggestions for fine-tuning of the parallel algorithms presented here. Additionally, W. Sands would like to thank the organizers of the 2019 Performance Portability Workshop with Kokkos, along with Christian Trott, David Hollman, and Damien Lebrun-Grandie for their assistance with our application. We also wish to express our gratitude to W. Nicolas G. Hitchon for taking the time to review our work and offer numerous suggestions and improvements regarding the presentation of the manuscript.

% SIAM bib format + include for the references
\medskip

\printbibliography

\end{document}